\documentclass[sigconf, nonacm]{acmart}

\usepackage{algorithm}
\usepackage{algorithmicx}
\usepackage{subfigure}
\usepackage[noend]{algpseudocode}
\usepackage{xcolor}
\usepackage{amsmath}
\usepackage{amsthm}
\usepackage{multicol}
\usepackage{balance}

\usepackage{multirow}
\usepackage{makecell}

\newtheorem{theorem}{Theorem}[section]

\newtheorem{definition}{Definition}[section]

\newcommand{\Paragraph}[1]{\smallskip\noindent{\bf #1.}}

\newcommand{\Removed}[1]{}

\newcommand{\techreport}[1]{#1}
\newcommand{\nottechreport}[1]{}

\definecolor{myblue}{rgb}{0,0,0.7}
\definecolor{mypurple}{rgb}{0.5,0,0.5}
\newcommand\vldbdoi{xx.xx/xxx.xx}

\newcommand\vldbavailabilityurl{https://github.com/schencoding/lits}

\begin{document}

%

\widowpenalty=0
\clubpenalty=0
\flushbottom


\title{LITS: An Optimized Learned Index for Strings \\ (An Extended
  Version)}

\author{Yifan Yang \ \ \ \ \ \ Shimin Chen}
\authornote{Shimin Chen is the corresponding author.}
\affiliation{%
  \institution{SKLP, ACS, Institute of Computing Technology, CAS}
  \institution{University of Chinese Academy of Sciences}
}
\email{{yangyifan22z, chensm}@ict.ac.cn}

\begin{abstract}

    Index is an important component in database systems.  Learned indexes
    have been shown to outperform traditional tree-based index structures
    for fixed-sized integer or floating point keys.  However, the
    application of the learned solution to variable-length string keys is
    under-researched.  Our experiments show that existing learned indexes
    for strings fail to outperform traditional string indexes, such as HOT
    and ART.  String keys are long and variable sized, and often contain
    skewed prefixes, which make the last-mile search expensive, and
    adversely impact the capability of learned models to capture the
    skewed distribution of string keys.

    In this paper, we propose a novel learned index for string keys, LITS
    (\underline{L}earned \underline{I}ndex with Hash-enhanced Prefix
    \underline{T}able and \underline{S}ub-tries).  We propose an optimized
    learned model, combining a global Hash-enhanced Prefix Table (HPT) and
    a per-node local linear model to better distinguish string keys.
    Moreover, LITS exploits compact leaf nodes and hybrid structures with
    a PMSS model for efficient point and range operations.
    Our experimental results using eleven string data sets show that LITS
    achieves up to 2.43x and 2.27x improvement over HOT and ART
    for point operations, and attains comparable scan performance.

\end{abstract}

\maketitle



\section{Introduction}
\label{sec:introduction}
\label{Re:R2O3}

Indexes play an essential role in modern database engines to
accelerate transaction and query processing.  Learned indexes have
been shown to outperform traditional tree-based index structures for
fixed-sized integer or floating point keys~\cite{LIPP, ALEX, PLEX,
    MADEX, RS, RMI, FINEdex, DILI, XIndex, LI_Ready}.  However, this is
hardly the case for variable-length string keys, which are
common in the real world~\cite{AtikogluXFJP12,CaoDVD20}.


While learned indexes have been extensively studied for fixed-sized
integer or floating point keys in recent years, the application of the
learned solution to variable-length string keys is under-researched
with only a couple of studies~\cite{SIndex,RSS}.
We experimentally compare the existing learned indexes for strings,
i.e. SIndex~\cite{SIndex} and RSS~\cite{RSS}, with state-of-the-art
traditional string indexes, i.e., ART~\cite{ART} and HOT~\cite{HOT}.
We find that existing learned indexes fail to outperform traditional
indexes.  In fact, traditional string indexes win by a large margin.

By examining real-world string data sets, we observe two distinct
features of string keys that differ significantly from fixed-sized
integer or floating point keys.  First, string keys are often
\emph{long and variable sized}, making the key access and comparison
more expensive.  Second, string data sets see \emph{skewed prefixes}
among string keys.  Popular prefixes shared by multiple strings make
it difficult for learned models to distinguish individual string keys.

The two distinct features impact the tree height, the node search, and
the last-mile search in index structures. For example, the last-mile
search often requires expensive key comparisons, and therefore should
be avoided as much as possible.  Recent studies attempt to adapt CDF
models for fixed sized keys (e.g.  RMI~\cite{RMI}, Radix
Spline~\cite{RS}, and piece-wise linear models) to string data sets.
However, the resulting learned models work poorly for capturing the
skewed distribution of string keys, leading to large tree heights that
degrade index performance.

In this paper, we propose a novel learned index for string keys, LITS
(\underline{L}earned \underline{I}ndex with Hash-enhanced Prefix
\underline{T}able and \underline{S}ub-tries).  First, LITS employs the
collision-driven design of LIPP~\cite{LIPP} to avoid the last-mile
search by creating a child node to store the keys that are mapped to
the same slot.
Second, we propose an optimized learned model, combining a global
Hash-enhanced Prefix Table (HPT) and a per-node local linear model.
The HPT approximates the conditional probability of the next character
given a prefix in the string key.  Compared with existing learned
models, HPT can better distinguish string keys.
Third, the collision-driven design can result in a large number of
small leaf nodes containing two or only a few keys.  Consequently, a
scan may have to traverse many small nodes, incurring expensive cache
misses and node jump overhead.  We introduce the compact leaf node,
which replaces a group of small nodes with a single node.
Finally, we observe that trie-based index, such as HOT, is very
efficient for highly skewed string data sets.  Therefore, we combine
our learned index and HOT using a performance model (PMSS) to
determine whether a subtrie is more beneficial to be used in the place
of a child node.



LITS supports common index operations on string keys,
including bulkload, search, insert, delete, update, and range scans.
It is specifically optimized for point operations. We conduct
extensive experiments using seven real-world string data sets
and four synthetic data sets.  Our experimental results show that LITS
achieves up to 2.06x and 2.14x improvement over HOT and ART for point
operations, respectively.  For the scan-heavy workload, LITS's
performance is comparable with HOT and better than ART.

\Paragraph{Contributions}
The contributions of the paper are as follows.
First, we propose a novel HPT-based CDF model that exhibits strong
discriminative power for string keys.  Second, we propose LITS, a
novel learned index for strings that exploits the HPT-based model,
compact leaf nodes, and hybrid structures with a PMSS model for
efficient point and range operations.  Finally, we perform extensive
experiments to compare our proposed LITS with five state-of-the-art
string indexes using eleven string data sets.  Our experiments show
that LITS achieves the overall best performance.

\Paragraph{Organization}
The rest of the paper is organized as follows.
Section~\ref{sec:background} studies the characteristics of string
keys and examine existing string indexes to motivate our study.  Then,
Section~\ref{sec:LITS} presents the LITS design.
Section~\ref{sec:experiment} experimentally compares LITS with
state-of-the-art string indexes.  Finally,
Section~\ref{sec:conclusion} concludes the paper.

\section{Background and Motivation}
\label{sec:background}

\begin{table}[t]

  \caption{String data sets used in this work. (cf.
    Section~\ref{subsec:setup})}
  \label{tab:datasets}
  \vspace{-0.15in}

  \centering
  \small
  \setlength{\tabcolsep}{3.5pt}

  \begin{tabular}{|l|c|c|c|c|c|}
    \hline
    Dataset        & Min Len & Max Len        & Avg Len & Number of Keys & Total Size \\
    \hline\hline
    \verb|address| & 4B      & 133B           & 24B     & 34M            & 802MB      \\ \hline
    \verb|dblp|    & 2B      & 255B$^\dagger$ & 76B     & 7M             & 506MB      \\ \hline
    \verb|geoname| & 2B      & 152B           & 13B     & 7M             & 106MB      \\ \hline
    \verb|imdb|    & 2B      & 106B           & 13B     & 9M             & 132MB      \\ \hline
    \verb|reddit|  & 3B      & 26B            & 11B     & 26M            & 292MB      \\ \hline
    \verb|url|     & 12B     & 255B$^\dagger$ & 64B     & 63M            & 4.6GB      \\ \hline
    \verb|wiki|    & 2B      & 255B$^\dagger$ & 15B     & 43M            & 870MB      \\ \hline
    \verb|email*|  & 11B     & 47B            & 23B     & 45M            & 1.1GB      \\ \hline
    \verb|idcard*| & 18B     & 18B            & 18B     & 63M            & 1.2GB      \\ \hline
    \verb|phone*|  & 11B     & 23B            & 17B     & 50M            & 819MB      \\ \hline
    \verb|rands*|  & 2B      & 61B            & 32B     & 50M            & 1.6GB      \\
    \hline
  \end{tabular}\\

  \raggedright
  \footnotesize
  \begin{list}{}{\setlength{\leftmargin}{4mm}\setlength{\itemindent}{-3mm}\setlength{\topsep}{0mm}\setlength{\itemsep}{0mm}\setlength{\parsep}{0mm}}

    \item $^\dagger$: The data set is processed to remove strings longer than 255B.  The maximum key length of the unprocessed \verb|dblp| is up to 1461B.

    \item *: The data set is synthetically generated.

  \end{list}

  \vspace{-0.16in}
\end{table}

\begin{figure}[t]
  \centering
  \includegraphics[height=3.5cm]{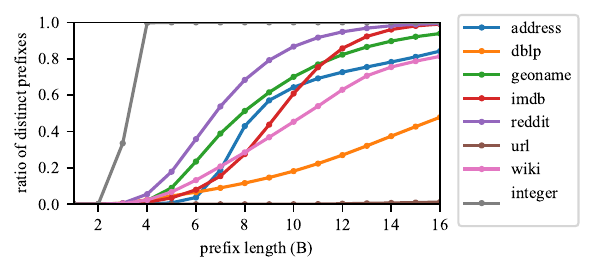}

  \vspace{-0.15in}

  \caption{Prefix skewness of string keys. }

  \label{fig:StringFeatures}
  \vspace{-0.20in}

\end{figure}

Learned indexes have been shown to outperform traditional tree-based
index structures for fixed-sized integer or floating point
keys~\cite{LIPP, ALEX, PLEX, MADEX, RS, RMI, FINEdex, DILI, XIndex, LI_Ready}.  However, this is hardly the case for string keys.  In
the following, we study the characteristics of string keys in
Section~\ref{subsec:string}, then examine existing index structures
optimized for strings to motivate our study in
Section~\ref{subsec:existing}.

\subsection{Characteristics of String Keys}
\label{subsec:string}

Table~\ref{tab:datasets} summarizes 7 real-world and 4 synthetic
string data sets used in this work.  (Please see detailed description
in Section~\ref{subsec:setup}.)  Focusing on the real-world data sets,
we observe two features that are distinct from fixed-sized integer or
floating point keys.

\Paragraph{Long and Variable Sized Keys}
Integer or floating point keys are typically of 4B or 8B large.  In
comparison, the string keys in the real-world data sets are much more
complex.  They can vary from 2B to over 1KB.  The average key length
of the real-world data sets is from 11B to 26B, which is much longer
than 4B/8B keys.
Consequently, storing entire string keys in index (inner) nodes can
significantly reduce node fanouts, degrading index performance.  On
the other hand, storing pointers to string keys in index nodes causes
pointer dereferences, incurring CPU cache misses.
Moreover, the comparison of long keys is also more expensive.

\Paragraph{Skewed Prefixes}
The prefixes of string keys are often quite skewed.
Figure~\ref{fig:StringFeatures} compares the prefix skewness of the
real-world string data sets and a uniformly generated integer data
set.   For each prefix length $k$, we compute the ratio of distinct
prefixes of a data set as the number of distinct k-byte prefixes
divided by the total number of keys in the data set.
This ratio is between 0 and 1.  If it is closer to 1, then the data
set is more evenly distributed.  If it is closer to 0, then a few
prefixes are very popular.  A large number of keys share the same
prefixes.  The data set is more skewed.
In Figure~\ref{fig:StringFeatures}, we consider the prefix length when
the ratio of a data set is over 0.99.  For the integer data set, all
keys can be distinguished by four bytes.  In contrast, all real-world
string data sets have very low ratios of distinct prefixes at 4B
prefixes.  For \verb|reddit|, the ratio reaches 0.99 at 16B prefixes.
The ratio of \verb|url| gets to 0.99 at 154B prefixes.  Consequently,
it is necessary to examine a much larger number of bytes for
distinguishing string keys, adversely impacting the effectiveness of
learned models in learned indexes.

\subsection{Existing Indexes Optimized for Strings}
\label{subsec:existing}

We focus on ordered indexes for strings in this paper.
Figure~\ref{fig:CI} compares the search performance of five
state-of-the-art index structures optimized for strings, including two
trie-based indexes (i.e., ART~\cite{ART} and HOT~\cite{HOT}), and
three learned index based structures (i.e., SIndex~\cite{SIndex},
RSS~\cite{RSS}, SLIPP, which is based on LIPP~\cite{LIPP})\footnote{
  We do not include B+-Trees in the comparison because previous work has
  shown that trie-based indexes significantly outperform B+-Trees for
  string data sets~\cite{HOT}. Recent work on Extendible Radix Tree
  (ERT) enhances trie nodes with extendible hashing~\cite{WhenTreeMeetHash}.  While the
  idea could potentially support string keys, the original paper and its
  code focus only on integer keys.  Therefore, we do not consider ERT in
  this work.
}.  From the figure, it is clear that existing learned indexes work
poorly compared to traditional trie-based indexes.  In the following,
we examine the index design choices to understand the pros and cons of
the existing index designs.

\begin{figure*}
  \centering
  \includegraphics[width=16cm]{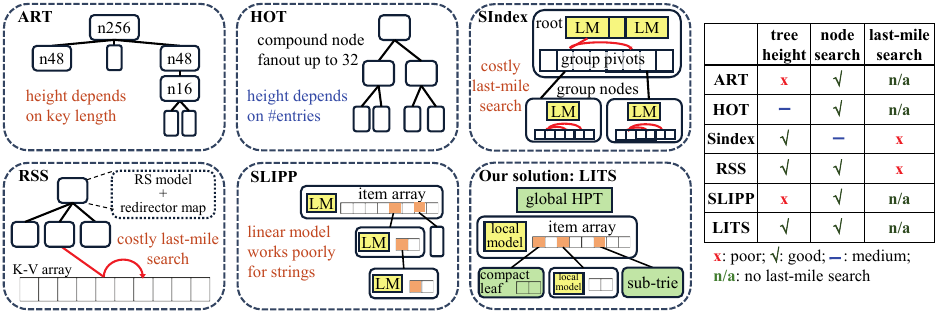}

  \vspace{-0.1in}
  \caption{Comparing index structures optimized for strings.}
  \label{fig:indexes}
  \vspace{-0.2in}

\end{figure*}


\Paragraph{Index Performance Factors}
Ordered indexes are typically organized as a tree.  All the five
state-of-the-art string indexes are essentially trees consisting of
inner nodes and leaf nodes.  A search often starts from the root of a
tree, visits several inner nodes at different tree levels, and finally
reaches a leaf node in the tree.  Therefore, the number of tree levels
from the root to the leaf and the search procedures at inner and leaf
nodes are main factors influencing the index search performance:

\begin{list}{\labelitemi}{\setlength{\leftmargin}{5mm}\setlength{\itemindent}{-1mm}\setlength{\topsep}{0.5mm}\setlength{\itemsep}{0.5mm}\setlength{\parsep}{0.5mm}}

  \item \emph{Tree height}:
        The (average) tree height indicates the expected number of nodes
        accessed by an index search.  Each node access often incurs an
        expensive CPU cache miss. These cache misses are dependent on each
        other since the memory address of the node at the next level is known
        only after searching the node at the previous level. Therefore, the
        tree height is an important performance factor.
        We see four cases for tree heights.  First, the tree height is
        determined by (the logarithm of) the number of index entries (e.g., in
        B+-Trees).
        Second, the tree height is determined by the length of the index keys
        (e.g., in ART or RSS).
        Third, learned indexes introduce CDF models to predict the key
        positions in a node in order to increase node sizes and reduce the
        tree height.  The tree height is \emph{model-based}.
        Finally, SIndex~\cite{SIndex} constructs a two-level tree with a root
        node and a number of group nodes.


  \item \emph{Node search}:
        Search in an inner node narrows down the search scope to a subtree of
        the node.  Search in a leaf node locates the target index entry.  Both
        often follow similar search procedures.  There are mainly three ways
        to support node search.
        First, it can be based on \emph{key comparisons} over the full keys or
        on part of the keys, e.g., binary search in a sorted B+Tree node.
        Second, it can perform an \emph{array lookup}, e.g., in a common trie
        node, such as node256 in ART.
        Finally, learned indexes often conduct \emph{model-based} search,
        which employs a learned model to predict the location of the search
        key in the key array.


  \item \emph{Last-mile search}:
        In learned indexes, model prediction in inner nodes is often accurate
        by construction.  Index keys are mapped to subtrees of an inner
        node using the associated model during bulkload and write operations.
        However, the models in leaf nodes may not predict the correct key positions.
        Therefore, learned indexes often have to do extra
        work, a.k.a. \emph{last-mile search}, to locate the key around the
        predicted position in leaf nodes.  The last-mile search often performs
        key comparisons (e.g., with exponential or binary search), and can
        incur poor performance for learned indexes with fixed-sized
        keys~\cite{LI_Ready}.  The situation is even worse for
        long and variable sized keys because of pointer dereferences and higher key
        comparison costs.


\end{list}

\begin{figure*}[t]
  \centering
  \includegraphics[width=16cm]{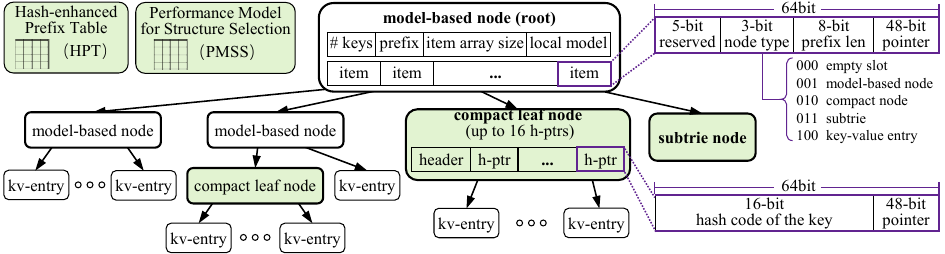}
  \vspace{-0.2in}
  \caption{Overview of LITS.}
  \label{fig:lits}
  \vspace{-0.2in}
\end{figure*}

\Paragraph{Pros and Cons of Existing Indexes}
We examine the three performance factors of the five string indexes.
Figure~\ref{fig:indexes} compares the five indexes and our proposed
solution.

\begin{list}{\labelitemi}{\setlength{\leftmargin}{5mm}\setlength{\itemindent}{-1mm}\setlength{\topsep}{0.5mm}\setlength{\itemsep}{0.5mm}\setlength{\parsep}{0.5mm}}

  \item \emph{ART}:
        Adaptive Radix Tree (ART)~\cite{ART} is a compressed trie.
        Each level of an ART uses a byte in the keys.  Therefore, the height
        of an ART is determined by the key lengths, which can be quite large
        for string data sets.
        ART compresses the trie node to four types of nodes (i.e.,
        Node4/16/48/256) to reflect the effective node fanouts.  The node
        search in Node48 and Node256 performs array lookups, while Node4 and
        Node16 employ key comparison based search.   Both procedures are fast
        because ART searches 1B in every node.  There is no last-mile search.

  \item \emph{HOT}:
        Height Optimized Trie (HOT)~\cite{HOT} optimizes ART by reducing the
        tree height.  Each inner node of a HOT is a compound node that
        represents a Patricia trie with a fanout of up to 32.  This is
        achieved by carefully storing only a subset of distinct key bits
        (a.k.a. partial keys) in each node.  As a result, HOT often reduces
        the tree height significantly for long string keys.  Its height can be
        viewed as roughly determined by the number of index entries.
        Moreover, the node search compares partial keys with efficient SIMD
        operations.  There is no last-mile search.

  \item \emph{SIndex}:
        SIndex~\cite{SIndex} is a two-level tree consisting of a root node and
        a level of group nodes.  The root node employs a piece-wise linear
        model (PLM) and divides the key space into key groups.  Then, each
        group node uses a linear model (LM) to locate an index entry.  Since
        the model prediction is not fully accurate, SIndex performs a
        last-mile search, which is a binary search within the error bound
        around the predicted location, in both the root and the group nodes.
        The last-mile search incurs significant performance overhead.

  \item \emph{RSS}:
        Radix String Spline (RSS)~\cite{RSS} is a trie.  Each trie node uses
        8B or 16B of the keys, and computes a Radix Spline (RS)
        model~\cite{RS} for them.  The RS model is a piece-wise linear model
        that provides monotonic CDF prediction with a given error bound.  For
        keys dissatisfying the error bound (e.g., because of shared prefixes),
        RSS stores the key in the redirector map and creates a new child node
        for the key.  RSS compares 8B/16B portions of keys, which has lower
        cost than full key comparison.  However, the last-mile search is still
        very costly. In the search experiments, RSS spends over 70\% of the
        time in the last-mile search.

  \item \label{Re:R5W3} \emph{SLIPP}. LIPP~\cite{LIPP} is an
        interesting learned index with fixed-sized keys because it avoids the
        costly last-mile search.
        In each inner node, LIPP trains a linear model.  If multiple
        keys are mapped to the same entry slot by the linear model, LIPP
        creates a new child node for the collision keys.
        This design is known as the collision-driven approach.  It essentially
        converts the last-mile search into a sub-tree search.
        We implement a variant of LIPP, called \emph{SLIPP}, to support string keys.
        At each node, SLIPP computes a numeric representation of each string
        key (after excluding the common prefix of the keys in the node) using
        a straight-forward formula: $y=\frac{s_1}{256} + \cdots  +
          \frac{s_{len}}{{256}^{len}}$, where $s_i$ is the $i$-th byte of the
        key.  Then, it computes the linear model based on the numeric
        representation.  The other design of LIPP is kept unchanged.
        While SLIPP avoids the last-mile search, the computed model can hardly
        distinguish keys with skewed prefixes, resulting in many collisions
        and large tree heights.



\end{list}

\Paragraph{Motivation}
The above discussion focuses on search, which is representative of
point operations.  The comparison of the
five string index structures motivate us to design an optimized
learned index for strings that avoids the last-mile search and
improves the effectiveness of the learned models for reducing the tree
height.


\section{LITS}
\label{sec:LITS}

We propose LITS (\underline{L}earned \underline{I}ndex with Hash-enhanced
Prefix \underline{T}able and \underline{S}ub-tries) for string keys in this
section.  Section~\ref{subsec:overview} overviews the structure and operations
of LITS.  Then, Section~\ref{subsec:HPT},~\ref{subsec:cnode},
and~\ref{subsec:SDT} present the three main techniques of LITS, optimizing the
learned model, accelerating scans with compact leaf nodes, and exploiting
subtries to further improve performance, respectively.  Finally,
Section~\ref{subsec:cost} describes the time and space cost of LITS.

\subsection{Overview of LITS}
\label{subsec:overview}

Figure~\ref{fig:lits} depicts the structures in LITS.  We describe the
distinct features of LITS (highlighted in the figure) in the following.

\begin{list}{\labelitemi}{\setlength{\leftmargin}{5mm}\setlength{\itemindent}{-1mm}\setlength{\topsep}{0.5mm}\setlength{\itemsep}{0.5mm}\setlength{\parsep}{0.5mm}}
      
      \item \emph{Model-based node}:
            To deal with the problem of the last-mile search, 
            as discussed in Section~\ref{subsec:existing}, 
            we employ the collision-driven design of LIPP
            to avoid the last-mile search.  Specifically, a model-based node
            consists of a header and an item array.  The header contains metadata,
            such as the number of keys, the key prefix, the size of the item
            array, and a local linear model.  Each slot in the item array is a
            64-bit pointer.  We store additional information in the upper bits of
            the pointers, which are otherwise unused in current machines.  Keys
            are mapped to slots in the array with an optimized learned model
            (discussed in more detail below).  There are three cases.  First, a
            slot is empty.  Then, it contains a NULL item.  Second, only a single
            key is mapped to a slot. Then, the slot holds a pointer to the
            key-value entry.  Third, multiple keys are mapped to the same slot.
            Then, LITS creates a child node to store the keys to avoid the
            last-mile search.  The node type field in the 64-bit item indicates
            the different child node types.
            
      \item \emph{Optimized global HPT and local models}:
            As discussed in Section~\ref{subsec:existing}, due to skewed prefixes
            and long keys, existing learned models work poorly for string keys.
            This results in the large tree height in SLIPP, lowering index
            performance.  \label{Re:R5D13} We propose an optimized learned model, 
            combining a global Hash-enhanced Prefix Table (HPT) and a per-node local 
            linear model to effectively distinguish string keys. (cf. Section~\ref{subsec:HPT})
            
      \item \emph{Compact leaf node}:
            The collision-driven design in model-based nodes lead to a large
            number of small leaf nodes that contain two or only a few kv-pointers.
            However, a scan has to traverse many such small leaf nodes, and
            suffers from expensive cache misses and node jump overhead.  We
            introduce the compact leaf node to make the design scan-friendly.  A
            compact node contains a header and an array of h-pointers sorted in
            the key order.  An h-pointer consists of a 16-bit computed hash of the
            key and a 48-bit pointer to the kv-entry.  In this way, we replace a
            number of small leaf nodes with a single compact leaf node, thereby
            reducing the number of node visits in scans.  (cf.
            Section~\ref{subsec:cnode})
            
      \item \emph{Subtrie node and PMSS}:
            We call the resulting index with the above techniques, LIT.  Our
            experiments show that LIT outperforms all the five existing indexes
            for most data sets, but it is slightly slower than the trie-based
            index (i.e., HOT) for a couple of data sets.  Therefore, we propose to
            combine LIT and trie-based indexes.  We build a performance model
            (i.e., PMSS) to determine whether a subtrie is more beneficial to be
            used in the place of a child node.  The combined structure of LIT with
            subtries is our final proposed solution, LITS. (cf.
            Section~\ref{subsec:SDT})
            
\end{list}

\noindent
After overviewing the structures of LITS, we describe the common index
operations in the following.

\Paragraph{Search}
A search goes from the root node to a leaf node.  Based on the node
type, LITS performs different search procedures.  

First, in a model-based node, LITS compares the common prefix recorded
in the node header with the search key.  Most commonly, the prefixes
match. Then, LITS skips the prefix and uses the remaining substring of
the search key to predict the slot position based on the global HPT
and the local model.  The position is between $1$ and
$ItemArraySize-2$.  In rare situations, the prefixes do not match.  We
preserve the first (the last) item of the item array for the case
where the search key prefix is less (greater) than the recorded
prefix.  Then, LITS gets the target item according to the search key.
If the item is NULL, the search key does not exist.  Otherwise, LITS
dereferences the pointer to visit the child node / kv-entry.

Second, in a compact leaf node, LITS performs a key comparison based
search.  It dereferences an h-pointer only if the hash of the search
key matches the hash in the h-pointer.

Third, if the node is a subtrie node, LITS calls the search procedure
of the subtrie (e.g., HOT) to continue the search.

Finally, upon reaching a kv-entry, LITS compares the search key with
the key in the kv-entry.  The search succeeds if it is a match.
Otherwise, the search key is not found.

\Paragraph{Insert/Delete/Update}
An insert first performs the search for the input key.  If the search
gets to an empty item, LITS simply inserts the new kv-entry to the
empty slot.  If the search gets to a single-entry item (i.e., a
pointer to a kv-entry), LITS builds a new compact leaf node to contain
the new key and the existing key.  If the search gets to a compact
leaf node, then LITS inserts the key to the compact leaf node if there
are less than 16 entries. \label{Re:R5D1}
When the compact node already contains 16 entries, LITS
performs the PMSS-based decision and replaces the compact node with
either a model-based node or a subtrie.
If the search gets to a subtrie, then LITS calls the insert procedure
of the subtrie to complete the insertion.

The delete or update procedures work similarly.  For an update, LITS
searches the key and either modifies the value in the kv-entry or
changes the item or h-pointer to point to a new kv-entry.  For a
delete, LITS removes the key if it is found.  This clears the
single-entry item, or reduces the number of keys in a compact node.
The situation is a little complicated for subtrie deletion because
HOT does not implement the delete function.  We employ a delete list
to hold the deleted keys associated with a subtrie.  If the number of
deleted keys is beyond a predefined ratio of the keys in the subtrie,
we reconstruct the subtrie.  

When a model-based node contains too many (or too few) keys, LITS
follows a procedure similar to LIPP to perform node resizing
operations.  Moreover, when inserting to a compact leaf node with 16
entries, LITS uses the PMSS to determine if a model-based node or a
subtrie should be constructed.  (Please see Section~\ref{subsec:SDT}
for all cases where PMSS-based decision is performed.)  Finally, we
adapt the classic method of optimistic locking, which protects each
node with a lock and allows reading a node without locking
it, to support concurrent threads.

\Paragraph{Scan}
Given a begin key, a scan searches the begin key and constructs an
iterator.  Using the iterator, one can obtain a list of kv-pointers
sorted by the key order by repeatedly calling the iterator's next
method.  Internally, the scan maintains a stack of pointers to nodes
from the root to the current leaf node.  Both the items in model-based
nodes and the h-pointers in compact leaf nodes are sorted in the key
order.  Therefore, the scan can easily traverse the nodes in the tree
using the stack.

\Paragraph{Bulkload}
At the beginning, LITS samples a subset of keys and compute the global
HPT model. 
Then, LITS bulkloads the tree in a similar fashion as LIPP.  There are
two main differences. First, for a sub range of data, LITS chooses
which node type to build.  If the number of keys is at most 16, then
LITS builds a compact node.  If there are more keys, LITS uses the
PMSS to choose between a model-based node and a subtrie.
Second, when constructing a model-based node, LITS uses the global HPT
to compute the local linear model for the keys in the node.

\subsection {Hash-enhanced Prefix Table (HPT)}
\label{subsec:HPT}

We would like to design a good learned model to better distinguish
string keys.  In the following, we begin by deriving a recursive
formula for CDF computation.  Using the formula, we explain why
previous linear models work poorly.  Then, we propose our HPT-based
model to better approximate the CDF.  Finally, we describe the training
and computation procedure using the HPT.

\Paragraph{Recursive Formula for CDF Computation}
Given a string data set and a string $S$=$s_1...s_{n}$, we would like
to compute $cdf(S)$.  For brevity of presentation, we prepend a
special (non-existent) beginning character $s_0$ to every string.
Hence, $S$=$s_0s_1...s_{n}$.  We denote the $k$-byte prefix of the
string as ${\mathcal{P}}_k$=$s_0s_1...s_{k}$.  Therefore,
$S\equiv{\mathcal{P}}_n$.  Then, we have the following recursive
formula:
\begin{equation}\label{eqn:cdf}
  \begin{aligned}
     & cdf({\mathcal{P}}_{0})    =  0                                                  \\
     & cdf({\mathcal{P}}_{k+1})  =  cdf({\mathcal{P}}_{k}) +
    prob({\mathcal{P}}_{k}) \times \Sigma_{c=0}^{s_{k+1}-1}{prob(c|{\mathcal{P}}_{k})} \\
  \end{aligned}
\end{equation}
Here, $prob({\mathcal{P}}_{k})$ represents the probability of prefix
${\mathcal{P}}_{k}$ in the string data set.
$prob(c|{\mathcal{P}}_{k})$ stands for the conditional probability of
the next character being $c$ given the prefix ${\mathcal{P}}_k$.

%
%
%
%


We can also derive a recursive formula for $prob({\mathcal{P}}_{k})$ as
follows:
\begin{equation}\label{eqn:prob}
  \begin{aligned}
     & prob({\mathcal{P}}_{0})   = 1                               \\
     & prob({\mathcal{P}}_{k+1}) = prob({\mathcal{P}}_{k}) \times 
    prob(s_{k+1}|{\mathcal{P}}_{k}) 
  \end{aligned}
\end{equation}
From Eqn~\ref{eqn:cdf} and~\ref{eqn:prob}, it is clear that obtaining
$prob(c|{\mathcal{P}}_{k})$ for any prefix ${\mathcal{P}}_{k}$ and any
character $c$ is crucial for computing $cdf(S)$.

\Paragraph{Problem of Existing Linear Models}
Existing linear models predict the position of a string
$S$=$s_1...s_{n}$ as a linear function: $y(S) = \alpha \times x +
  \beta$, where $x$ = $\Sigma_{k=1}^{m}{\tfrac{s_k}{256^k}}$.   SLIPP
computes $x$ based on the full string, and hence $m$=$n$.  RSS uses an
8B or 16B portion of the keys in each node in the model prediction.
Therefore, $m$= 8 or 16 in RSS.
We can rewrite the formula in a recursive fashion as follows:
\begin{equation}\label{eqn:lm}
  \begin{aligned}
    y({\mathcal{P}}_{k+1}) = y({\mathcal{P}}_{k}) 
    + \tfrac{\alpha}{256^k} \times \tfrac{s_{k+1}}{256}
  \end{aligned}
\end{equation}
Since $y(S)$ is a scaled version of $cdf(S)$, we can compare
Eqn~\ref{eqn:cdf} and~\ref{eqn:lm}.  $\tfrac{\alpha}{256^k}$
corresponds to a scaled version of $prob({\mathcal{P}}_{k})$, and
$\tfrac{s_{k+1}}{256}$ corresponds to
$\Sigma_{c=0}^{s_{k+1}-1}{prob(c|{\mathcal{P}}_{k})}$.  
\label{Re:R5D2} $\tfrac{s_{k+1}}{256}$ implies that any 8-bit
character appears uniformly at random.  Therefore, the existing linear
models essentially assume that the distribution of the next character
following any given prefix is uniform.
This estimation can hardly reflect the true distribution in a string
data set, which often contains highly skewed prefixes.

%

%
%
%

\begin{figure*}
  \centering
  \includegraphics[width=14.5cm]{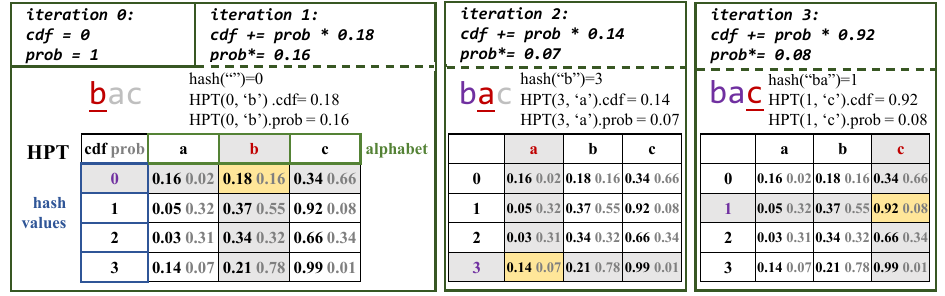}
  
  \vspace{-0.15in}
  \caption{An illustration of the CDF computation using the HPT for
    string ``bac''. (purple: prefix; red: current character)}
  \label{fig:HPT}
  \vspace{-0.15in}
  
\end{figure*}

\Paragraph{Our Solution: HPT}
We would like to better approximate the conditional probability
$prob(c|{\mathcal{P}})$.  Existing neural network-based nonlinear CDF
models~\cite{nonlinearLearnedIndex1, nonlinearLearnedIndex2} are more
accurate than linear models.  However, these models are complex.  The
model training and model prediction are time consuming.  It would be
an over-kill for our goal of designing efficient string indexes.

A na\"ive idea is to record the conditional probabilities for all
possible (prefix, character) pairs in the string data set.   However,
such an approach would require prohibitively large space to store the
conditional probabilities.

To reduce the space overhead, we propose the Hash-enhanced Prefix
Table (HPT).  As illustrated in Figure~\ref{fig:HPT}, the HPT is a
table (i.e., 2D array).  For any prefix, we map the prefix to a row in
the table using a hash function.  (We set the hash of the empty
prefix, $hash(s_0)$=0.) Then, each column corresponds to a character
in the character set. We approximate the conditional
probability $prob(c|{\mathcal{P}})$ with table lookups as
HPT[$hash({\mathcal{P}})$][$c+1$].cdf - HPT[$hash({\mathcal{P}})$][$c$].cdf.  We have stored this value in the model as HPT[$hash({\mathcal{P}})$][$c$].prob.
Note that HPT[$hash({\mathcal{P}})$][$c$].cdf approximates
$cdf(c|\mathcal{P})$, which is $\Sigma_{0}^{c-1}{prob(i|{\mathcal{P}})}$. This
reduces the complexity for the CDF computation.

\Paragraph{HPT Construction}
The construction of the HPT is simple. We randomly sample a small
fraction (e.g., 1\%) of the string data set during bulkloading, and
compute the HPT using the sample.
First, we initialize the HPT table to all 0s.
Second, we iterate through all the string keys in the sample.  For
each string, we extract all (prefix $\mathcal{P}$, character $c$)
pairs, and increment the corresponding cell
HPT[$hash({\mathcal{P}})$][$c$].  After the processing, each cell
contains the frequency of ($hash({\mathcal{P}})$, $c$).  
Finally, we process each row in the HPT.  We compute the accumulate
frequencies, and divide them by the total frequencies in the row to
obtain $cdf(c|hash(\mathcal{P}))$.  We store
$cdf(c|hash(\mathcal{P}))$ in HPT[$hash({\mathcal{P}})$][$c$].cdf and $cdf(c+1|hash(\mathcal{P}))$ - $cdf(c|hash(\mathcal{P}))$ in HPT[$hash({\mathcal{P}})$][$c$].prob. 


\addtolength{\textfloatsep}{-0.3in}
\begin{algorithm}[t]
  \caption{HPT-based CDF computation.}
  \label{alg:HPT_GetCDF}
  \begin{algorithmic}[1]
    \Procedure{GetCDF}{HPT, string S}
    
    \State cdf = 0, prob = 1
    
    \For{$k=0$ to $len(S) - 1$}
    \State hashval = (k == 0 ? 0 : hash(S[0:k-1]))
    \State c = S[k]
    \State cdf += prob * HPT[hashval][c].cdf
    \State prob *= HPT[hashval][c].prob
    \EndFor
    \State \textbf{return} cdf
    \EndProcedure
  \end{algorithmic}
\end{algorithm}
\addtolength{\textfloatsep}{0.3in}


\Paragraph{Model Prediction}
Algorithm~\ref{alg:HPT_GetCDF} shows the computation of $cdf(S)$ using
the HPT.  The initialization in Line 2 corresponds to
$cdf({\mathcal{P}}_{0})$ and $prob({\mathcal{P}}_{0})$.  Then, we use
Eqn~\ref{eqn:cdf} (Line 6) and Eqn~\ref{eqn:prob} (Line 7) to
iteratively compute the CDF and probability of the current prefix,
respectively.  To reduce the cost of the hash computation in Line 4,
we keep an internal state and incrementally update the state with the
next character in the string.  Then, we can compute the hash value of
the prefix with $O(1)$ cost.  The loop proceeds until the CDF of the
string S is computed.  Figure~\ref{fig:HPT} shows an example
computation of $cdf(bac)$.  

LITS combines the global HPT and the per-node linear model in model
prediction.  In a model-based node, the predicted position for string
$S$ is computed as $y(S) = \alpha \times x + \beta$, where $x$ =
$GetCDF$(HPT, $S$).  Note that we exclude the common prefix in this
computation.

\Paragraph{Benefits of the HPT-Based Model}
First, compared to the uniform assumption in the existing linear
models, our HPT-based model better captures the distribution of the
string data set.  Therefore, it can distinguish string keys more
effectively.
Second, the hashing design in the HPT reduces the space overhead for
recording the conditional probability distributions.  One can adjust
the number of HPT rows to balance the space cost and the estimation
quality.  The larger the HPT table, the higher the estimation quality.
However, a very large HPT table not only causes significant space
overhead, but also incurs random memory accesses and CPU cache misses
for HPT lookups.  Therefore, we set the HPT table size (e.g., 2MB in
our experiments) to be small enough to fit in the CPU cache.
Finally, our design is computationally efficient.  The HPT
construction using a sample of string keys is simple and fast.
Storing the conditional CDFs in the HPT reduces the cost for computing
the sum term in Eqn~\ref{eqn:cdf}.  Hence, model prediction takes
$O(len(S))$ time.



\Paragraph{Analysis of HPT Accuracy} \label{Re:R5W4}
We have the following theorem for the accuracy of 
approximating the conditional probability.
(Please refer to \nottechreport{the extended version of the
  paper~\cite{litstr}} \techreport{Appendix~\ref{app:hptproof}} for the
proof.)

\vspace{-1mm}
\begin{theorem} \label{thm:hpt}
  If prefix ${\mathcal{P}}$ appears $n_{{\mathcal{P}}}$ times in the
  string data set, and the HPT[$hash({\mathcal{P}})$] row sees $d$
  occurrences of other prefixes, then 
  \vspace{-1mm}
  $$|HPT[hash({\mathcal{P}})][c].prob - prob(c|{\mathcal{P}})| \leq
    \tfrac{1}{\tfrac{n_{{\mathcal{P}}}}{d}+1}.$$
  \vspace{-4mm}
\end{theorem}

\noindent
For a popular prefix ${\mathcal{P}}$, we expect $n_{{\mathcal{P}}} \gg
  d$ with a reasonable sized HPT (e.g., 2MB).  In such cases, the
absolute error of the HPT approximation is small.  
Our experiments confirm this result.  For the string data sets in our
experiments, the average absolute error of the conditional probability
is 0.0006--0.006 for popular prefixes that appear at least 10,000
times.


\Paragraph{Dealing with Data Distribution Changes}
\label{R4W2}
If the data distribution changes, HPT may become less accurate,
leading to degraded index performance.  To handle data changes, LITS
can sample the index performance (e.g., for 1\% of the queries).  If
it observes that the index performance falls below a pre-defined water
mark (e.g., 50\% of the average performance after bulkloading), LITS
can judiciously retrain the HPT model and rebuild the entire index.

\subsection{Compact Leaf Node}
\label{subsec:cnode}

\begin {figure}[t]
\centering
\includegraphics[width=8cm]{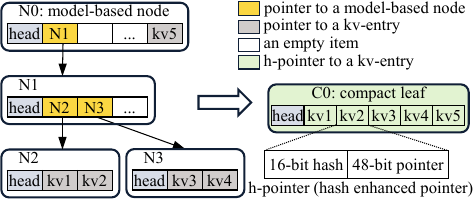}
\vspace{-0.1in}
\caption{Replacing multiple nodes with a compact leaf node.}
\label{fig:DeepBranch}
\vspace{-0.3in}
\end {figure}

The collision-drive design in the model-based nodes avoids the last-mile search
by creating new child nodes.  However, we observe that it can result in small
nodes with only two or a few keys, as illustrated in
Figure~\ref{fig:DeepBranch}.  The figure depicts a subtree with four
model-based nodes, i.e., $N0$--$N3$.  The four nodes are in three different
tree levels, while the entire subtree rooted at $N0$ contains only five
kv-entries.  This subtree structure is sub-optimal for the following reasons.
First, it degrades scan performance.  Suppose $kv1$--$kv5$ are retrieved by a
scan operation.  Then the scan has to traverse four nodes in three levels,
incurring expensive CPU cache misses and significant book-keeping overhead for
entering nodes and backtracing.
Second, the small nodes tend to increase the tree height, adversely impacting
point operations.  As shown in the example, $kv1$--$kv4$ are located two levels
deeper than $kv5$.  A search for $kv1$ is more costly than $kv5$.
Finally, the small nodes increase the space cost for storing the per-node
headers and the child node pointers.

To address this problem, we replace a number of small nodes with a single
compact leaf node.  As illustrated in Figure~\ref{fig:DeepBranch}, the four
nodes are replaced with a compact leaf node, holding the five kv-entries.
Each h-pointer stores a 16-bit hash of the key for better search performance.
A search in a compact node sequentially compares the hash of the search key
with the hash in every h-pointer.  Only when there is a match does LITS
dereference the pointer to visit the kv-entry.  The false positive rate with
the 16-bit hash is 0.0015\%. Compared to the common binary search, the
h-pointer based search can effectively avoid the high cost of unnecessary
kv-entry dereference and key comparison.
Moreover, the h-pointer array is sorted in the key order so that the scan
iterator avoids the cost of sorting the keys in the compact node.


We discuss two important design choices of the compact node.
\begin{list}{\labelitemi}{\setlength{\leftmargin}{5mm}\setlength{\itemindent}{-1mm}\setlength{\topsep}{0.5mm}\setlength{\itemsep}{0.5mm}\setlength{\parsep}{0.5mm}}
      
      \item \emph{Size threshold $w$ of compact nodes}:
            A compact node can hold up to $w$ keys.  If $w$ is too small, compact nodes may
            not effectively reduce the small nodes in the index.
            On the other hand, if $w$ is too large, the search performance suffers because
            it takes $O(w)$ time to sequentially examine the h-pointers.
            In Section~\ref{sec:experiment}, we study the impact of $w$ on LITS performance
            and set $w$=16 based on the experiments.

      \item \emph{Method to support inserts}:
            We consider two methods to support inserts.
            First, each compact node contains an array of $w$ slots.  If there are $k$
            keys, then $w-k$ slots are empty.  An insert operation places the new key into
            the existing array.  It moves existing elements to keep the sort order.
            Second, an alternative method is to make the node compact.  A compact node with
            $k$ keys is stored in an array of $k$ slots.  No space is reserved for empty
            slots.  Then an insert operation creates a new compact node with one more slot
            to hold both the existing keys and the new key.
            Our experiments find that the first method sees substantial space waste
            because of the reserved empty slots, and both methods have similar performance. 
            Therefore, we choose the second method as the default design for the compact node.
            
\end{list}

\subsection{LIT Enhanced with Subtries}
\label{subsec:SDT}

We call the learned index using HPT-based models and compact nodes,
LIT (\underline{L}earned \underline{I}ndex with Hash-enhanced Prefix
\underline{T}able).  In the following, we present a hardness metric,
GPKL, for string data sets.  We compare LIT and trie-based indexes
experimentally, and combine LIT with HOT using a GPKL-based
performance model to further improve index performance.

\Paragraph{Hardness of String Data Sets}
Previous study on learned indexes defined a hardness metric for data
sets with integer or floating point keys~\cite{LI_Ready}.  The metric
reflects the difficulty of applying linear models to approximate the
CDF of the data set.  However, this metric cannot be directly applied
to string data sets because linear models hardly capture the
properties of string data sets, as shown in Section~\ref{subsec:HPT}.
In this work, we propose a new hardness metric, GPKL (Group Partial
Key Length), for strings.



\begin{definition}[Common Prefix Length]
  
  The common prefix length of a list ${\mathcal{L}}$ of strings, denoted
  as $cpl({\mathcal{L}})$, is the length of the longest prefix shared by
  all strings in ${\mathcal{L}}$.
  
  \label{dfn:cpl}
\end{definition}

%
%

\begin{definition}[Partial Key Length]
  
  Given a sorted list ${\mathcal{L}}$ of strings, the partial key of the
  $i$-th string $S_i$ in ${\mathcal{L}}$ is the shortest substring of
  $S_i$ that distinguishes $S_i$ from $S_{i-1}$ and $S_{i+1}$ after
  removing the common prefix of ${\mathcal{L}}$.  The partial key length
  of $S_i$, denoted as $pkl({\mathcal{L}},S_i)$, is the length of
  $S_i$'s partial key.
  
  \label{dfn:dpl2}
\end{definition}

\noindent
$pkl({\mathcal{L}},S_i)$ can be computed with common prefix lengths as follows:
\begin{equation}\label{eqn:pkl1}
  pkl({\mathcal{L}}, S_i) = max(cpl(\{S_{i-1}, S_i\}), cpl(\{S_i, S_{i+1}\})) + 1 - cpl({\mathcal{L}})
\end{equation}
$cpl(\{S_{a}, S_b\})+1$ gives the smallest prefix length to distinguish 
$S_a$ and $S_b$.  Hence, the max term plus 1 shows the length of the
shortest prefix that distinguishes $S_i$ from $S_{i-1}$ and $S_{i+1}$.
Then, $pkl({\mathcal{L}}, S_i)$ is obtained by subtracting the common prefix
length of all strings in ${\mathcal{L}}$ from this shortest prefix length.

%
%

\begin{definition}[Group Partial Key Length]
  
  The group partial key length (GPKL) of a sorted list ${\mathcal{L}}$ of strings is
  the average of the partial key lengths of strings in ${\mathcal{L}}$: $gpkl({\mathcal{L}}) =
    \tfrac{1}{|{\mathcal{L}}|}\sum_{S \in {\mathcal{L}}}{pkl({\mathcal{L}}, S)}$.
  
  \label{dfn:dpl1}
\end{definition}

We choose GPKL as the hardness metric for strings for the following
reasons.
First, GPKL measures the difficulty of distinguishing keys in a string
data set.  The larger the GPKL, the more key bytes are necessary to
distinguish the strings.  Therefore, the metric reflects the hardness
of modeling the string data set.
Second, GPKL skips the common prefix of strings.  This behavior mimics
the design of inner nodes in most string indexes, including HOT, ART,
Sindex, RSS, and LIT.
Finally, GPKL can be computed efficiently by reading the sorted list
of strings in one pass.   This makes it possible to compute the GPKL
online for structure selection decisions.

We define both a global GPKL and a local GPKL metric.  Given a sorted
list of strings, the global GPKL is the GPKL of the entire list.  To
compute the local GPKL, we divide the sorted list into disjoint
sublists containing $g$ consecutive strings in the list.  We obtain
the GPKL for each sublist, then compute the average of the sublist
GPKLs as the local GPKL.  We set $g=32$ in the following.

\label{Re:R5D5}

\begin{table}[t]
  \caption{Impact of hardness on index performance (Mops).}
  \label{tab:LITPerf}
  \vspace{-0.15in}
  
  \centering
  \small
  \setlength{\tabcolsep}{4.2pt}
  
  \begin{tabular}{|l|c|c|ccc|ccc|}
    \hline
    String         & Global & Local & \multicolumn{3}{|c|}{Read-Only} & \multicolumn{3}{|c|}{Write-Only}                                               \\
    Dataset        & GPKL   & GPKL  & LIT                             & HOT                              & ART  & LIT           & HOT           & ART  \\
    \hline
    \verb|rands*|  & 6.12   & 2.42  & \textbf{3.37}                   & 2.62                             & 3.24 & \textbf{2.41} & 1.03          & 1.47 \\
    \verb|reddit|  & 8.24   & 3.48  & \textbf{3.01}                   & 1.90                             & 2.39 & \textbf{1.74} & 1.17          & 1.52 \\
    \verb|geoname| & 10.36  & 4.75  & \textbf{2.88}                   & 2.27                             & 2.27 & \textbf{1.62} & 1.26          & 1.45 \\
    \verb|imdb|    & 10.51  & 3.79  & \textbf{2.63}                   & 2.00                             & 1.97 & \textbf{1.88} & 1.23          & 1.35 \\
    \verb|phone*|  & 10.84  & 4.01  & \textbf{2.92}                   & 2.01                             & 2.38 & \textbf{1.53} & 1.18          & 1.43 \\
    \verb|address| & 12.61  & 6.55  & \textbf{2.23}                   & 2.08                             & 1.83 & \textbf{1.52} & 0.94          & 1.19 \\
    \verb|idcard*| & 12.89  & 5.04  & \textbf{3.19}                   & 1.92                             & 1.62 & \textbf{2.01} & 1.03          & 1.02 \\
    \verb|wiki|    & 14.32  & 6.23  & \textbf{1.94}                   & 1.68                             & 1.36 & \textbf{1.17} & 0.98          & 1.10 \\
    \verb|email*|  & 15.32  & 5.86  & 1.88                            & \textbf{1.89}                    & 1.11 & \textbf{1.06} & 0.92          & 1.00 \\
    \verb|dblp|    & 20.79  & 10.19 & 1.55                            & \textbf{1.93}                    & 1.30 & \textbf{0.88} & 0.72          & 0.83 \\
    \verb|url|     & 47.61  & 17.79 & 0.83                            & \textbf{1.27}                    & 0.78 & 0.54          & \textbf{0.68} & 0.58 \\
    \hline
  \end{tabular}
  \\note: * indicates that the data set is synthetically generated. 
  
  \vspace{-0.20in}
\end{table}

\Paragraph{Impact of Hardness on Index Performance}
HOT and ART significantly outperform existing learned indexes for
strings, as shown in Section~\ref{subsec:existing}.  Hence, we are
interested in comparing LIT with HOT and ART.  Table~\ref{tab:LITPerf}
reports the index throughput for both a read-only workload and a
write-only workload.  For the read-only workload, we randomly search
20 million keys after bulkloading an index with all keys.  For the
write-only workload, we bulkload an index with 50\% of the keys, and
then we measure the throughput of randomly inserting the rest of the
keys into the index.  

As shown in Table~\ref{tab:LITPerf}, the data sets are arranged in the
order of increasing global GPKLs.  We see that LIT achieves the best
read throughput for 8 data sets and the best write throughput for 10
out of the 11 data sets.  However, for the datasets with the highest
hardness values, trie-based indexes have higher performance.
Specifically, HOT has the best read performance for \verb|email|,
\verb|dblp|, and \verb|url|, and the best write performance for
\verb|url|.  This finding motivates us to combine the strengths of LIT
and HOT.

\Paragraph{Performance Model for Structure Selection (PMSS)}
To combine LIT and HOT, our basic idea is to make a decision to choose
from either LIT or HOT when creating a node for a subset of string
keys.  Obviously, it would be too costly to experimentally compare the
two choices online.  Therefore, we develop a performance model (PMSS)
to make quick and accurate online decisions.

\begin{figure}[t]
  \centering
  \includegraphics[width=8.5cm]{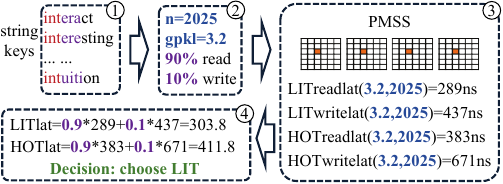}
  
  \vspace{-0.12in}
  \caption{The decision process with PMSS.}
  \label{fig:SDT_process}
  \vspace{-0.2in}
  
\end{figure}

The PMSS model works as follows.  We choose GPKL and the number ($n$)
of strings as two important metrics to characterize a subset of
strings.  For a given index, the PMSS model provides two functions,
$readlat$($gpkl$, $n$) and $writelat$($gpkl$, $n$), which estimate the
index search latency and the index insert latency, respectively.  A
target workload is specified as containing $f_r$ fraction of reads and
$f_w$ fraction of writes, where $f_r+f_w=1$.  (Operation statistics
can be updated online to estimate the $f_r$/$f_w$ parameters.) Then,
we estimate the average latency of index operations in the target
workload as follows:
\begin{equation}\label{eqn:pkl2}
  latency = f_r \cdot readlat(gpkl, n) + f_w \cdot writelat(gpkl, n) 
\end{equation}
We estimate the latency for each index, and select the design with the
lowest latency for the given subset of strings.
Figure~\ref{fig:SDT_process} illustrates this decision process using
the PMSS.

To obtain $readlat$($gpkl$, $n$) and $writelat$($gpkl$, $n$), we
perform a set of offline benchmarking tests using synthetically
generated data for various combinations of $gpkl$ and $n$, and
populate a $readlat$ table and a $writelat$ table for each index
(i.e., LIT and HOT).  In our experiments, we populate the tables for
$gpkl$=3, 5, ..., 21, and $n$= $2^4$, $2^5$, ..., $2^{25}$.  The total
size of latency tables for LIT and HOT is less than 10KB.  Then, for a
specific ($gpkl$, $n$), we can use the latency tables to easily
compute $readlat$($gpkl$, $n$) and $writelat$($gpkl$, $n$).

\begin{figure}[t]
  \centering
  
  \includegraphics[width=8cm]{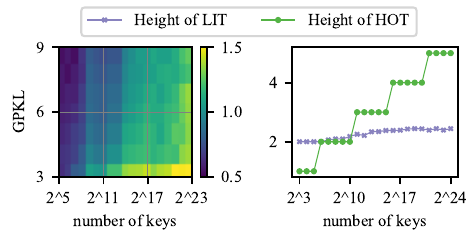}\\
  
  \vspace{-2mm}
  \hspace{1mm}
  \mbox{(a) speedup of LIT over HOT}
  \hspace{4mm}
  \mbox{(b) tree height (GPKL=5)}\\
  
  \vspace{-3mm}
  \caption{Offline benchmark tests for read-only workload.}
  
  \label{fig:SDTs}
  \vspace{-0.25in}
\end{figure}

Figure~\ref{fig:SDTs} displays the results of offline benchmarking
tests for the read-only workload.  Figure~\ref{fig:SDTs}(a) shows a
heat map.  We divide the $readlat$($gpkl$, $n$) of HOT by that of LIT
and then use different colors to represent the speedup.  The darker
color shows where HOT wins, while the brighter color shows where LIT
wins.  We see that for a fixed $gpkl$, LIT exhibits a leading
advantage as the data size increases.  This can be explained by
Figure~\ref{fig:SDTs}(b).  As the number ($n$) of keys increases, the
height of HOT increases significantly, roughly following
$\log_{32}(n)$, while the height of LIT only changes slightly.  As a
result, LIT outperforms HOT.


One interesting detail is how to generate a synthetic string data set
with specific $gpkl$=$l$ and $n$.  
First, we generate a random dictionary to contain 10000 random strings
that are 2B--6B long.  The strings are used as prefixes.
Second, we generate a set ${\mathcal{L}}$ of $n$ random strings, sort
the strings, and compute the initial $gpkl_0$ for ${\mathcal{L}}$,
which is typically small for randomly generated strings.
Third, we increase the $gpkl$ of ${\mathcal{L}}$ as follows.  We
randomly select $k$ adjacent strings $S_{a1}, S_{a2}, ..., S_{ak}$ in
the sorted list, and compute the common prefix length $cpl$ for the
$k$ strings.  Then, we randomly pick a string $S_p$ from the
dictionary, generate a random insert position $j$ $\in$ [0, $cpl$],
and insert $S_p$ into each $S_{ai}$ at the $j$-th byte.  In this way,
the $gpkl$ of ${\mathcal{L}}$ increases by at most $\tfrac{k \cdot
  len(S_p)}{|{\mathcal{L}}|}$.  We adjust the location of the $k$
strings to keep the sort order of ${\mathcal{L}}$.
Finally, we repeat the third step until the $gpkl$ reaches the target
$l$.  


\Paragraph{Structure Selection Scenarios}
LITS performs PMSS-based decisions in three main scenarios:
1) Bulkload: if a node corresponds to over 16 kv-entries, LITS uses
PMSS to decide whether to build a subtrie or a model-based node in the
bulkload operation;
2) Insert into a full compact node: When an insert sees a full compact
node (with 16 keys), it replaces the compact node with either a
model-based node or a subtrie;
3) Resize a model-based node: Like LIPP, LITS performs node resizing
if there are too many or too few keys in a model-based node $N_r$.
The resizing process rebuilds the subtree rooted at node $N_r$, and
uses the PMSS to decide whether a model-based node or a subtrie should
be constructed.

Moreover, LITS detects the case where over 50\% of the keys
are mapped to an index slot in a model-based node.  In such a case,
LITS builds a subtrie for the child node corresponding to the index
slot.  This restriction ensures that the non-subtrie part of the tree
is at most $O(logN)$ high.  Note that the 50\% restriction is actually
quite weak, and it has not been triggered in our experiments.




\Paragraph{Implementation Consideration}
It should be noted that careful design is required at the connection
point of different structures to avoid potentially unnecessary cost.
For example, the root node of HOT contains a single pointer to the
actual first level node.  Therefore, when creating a HOT subtrie, we
do not simply set the child pointer in the item to point to the root
node of HOT.  Instead, we directly replace that item with the root
node of HOT (while also handling the flag bits of both LIT and HOT
correctly).  In this way, we save a pointer dereference for accessing
the HOT subtrie.

\subsection{Cost Analysis}
\label{subsec:cost}
\label{Re:R2O2}

The time and space cost of tries (e.g., ART~\cite{ART} and
HOT~\cite{HOT}) have been extensively measured and studied.  In the
following, we mainly summarize the time and space cost of the
non-subtrie part of LITS.  (Please see the full description in
\techreport{Appendix~\ref{app:complexity}}
\nottechreport{the extended version of the paper~\cite{litstr}}.)
Suppose there are $N$ keys in the index.  Then, in the worst case, (1)
the height of the non-subtrie part of LITS is $O(logN)$, (2) the
search/update cost is $O(logN)$, (3) the amortized insert/delete cost
is $O({log}^2N)$, and (4) the space cost is $O(NlogN)$.

\section{Evaluation}
\label{sec:experiment}

In this section, we compare the performance of LITS with
state-of-the-art string indexes, and study the performance benefits of
our proposed techniques in LITS.

\subsection{Experimental Setup}
\label{subsec:setup}

\Paragraph{Machine Configuration}
All experiments are conducted on a machine equipped with two 3.4GHz
Intel Xeon Platinum 8380 CPUs (with 40 cores / 80 threads per CPU and
60MB L3 cache) and 256GB memory.  The machine runs the standard Ubuntu
20.04 Linux.  All programs are compiled with GCC 9.4.0 using the O3
optimization level.  To avoid NUMA effects, we perform the experiments
using a single CPU in the machine.

\Paragraph{Solutions to Compare}
We compare LITS with five state-of-the-art traditional and learned
indexes for strings:
\begin{list}{\labelitemi}{\setlength{\leftmargin}{5mm}\setlength{\itemindent}{-1mm}\setlength{\topsep}{0.5mm}\setlength{\itemsep}{0.5mm}\setlength{\parsep}{0.5mm}}

	\item \emph{ART (Adaptive Radix Tree)}~\cite{ART}:
	      We find
	      multiple ART implementations on github, and choose the one with the
	      highest stars (\url{https://github.com/armon/libart.git}).  The
	      implementation does not support scans.  Therefore, we add a scan
	      procedure for ART.

	\item \emph{HOT (Height Optimized Trie)}~\cite{HOT}:
	      We obtain the code written by the HOT authors
	      (\url{https://github.com/speedskater/hot.git}).

	\item \emph{SIndex}~\cite{SIndex}:
	      We use the implementation provided by the authors
	      (\url{https://github.com/curtis-sun/TLI.git}).
	      SIndex requires all strings to be padded to a uniform length.
	      Therefore, for SIndex experiments, we pad the strings in each data set
	      to the length of the longest string in the data set.

	\item \emph{RSS (Radix String Spline)}~\cite{RSS}:
	      For RSS, we cannot find publicly available code and write our own
	      implementation in C++.  We follow the RSS paper to employ the two-gram
	      compression of HOPE~\cite{HOPE} to encode string keys.  This improves
	      RSS's search performance.  The reported index performance includes
	      both the encoding of the query key and the actual index operation in
	      RSS.
	      However, RSS does not support insertions because it stores sorted key-value
	      data in an array, and uses the array indexes to indicate the key
	      ranges in tree nodes.  An insert would have to change the array and
	      update key ranges in tree nodes, which would be very costly.
	      Therefore, we omit RSS for all experiments that perform insertions.

	\item \emph{SLIPP}:
	      We obtain the LIPP~\cite{LIPP} code provided the LIPP
	      authors (\url{https://github.com/Jiacheng-WU/lipp}).  Then, we modify
	      LIPP to support strings as described in Section~\ref{subsec:existing}.
	      We implement the bulkload and the search operations.  We find
	      that SLIPP has much worse search performance than HOT, ART, and RSS.
	      Since SLIPP is clearly less competitive, we choose not to implement
	      the other operations for SLIPP and omit SLIPP for the rest of the
	      experiments. The implementation is written in C++.


	\item \emph{LITS}:
	      We implement LITS in C++.  The size of HPT is 2MB (with 1024 rows,
	      128 columns, and 16B per cell). A compact leaf node has a maximum
	      capacity of 16 elements.  The index used for constructing a hybrid
	      structure with LIT is HOT, because HOT demonstrates better overall
	      performance compared to ART.

	\item \emph{Variants of LITS}:
	      To understand the benefit of our proposed techniques, we also
	      implement several variants of LITS.  LIT is the learned index without
	      subtries.  Moreover, we change the learned model in LIT and implement
	      several LIT(model) variants, as will be described in
	      Section~\ref{subsec:HPTPerf}.  Furthermore, we study the combination
	      of LIT with different trie indexes.  LITS-A is LIT enhanced with ART
	      as the subtrie of choice.  (LITS-H is LIT enhanced with HOT, which is
	      another name for LITS.)

\end{list}

\noindent
All experiments are conducted using a single thread except for the
scalability experiments in Section~\ref{subsec:scalability}.  For the
scalability experiments, we compare LITS with the most competitive
solution, HOT.  The HOT code supports multiple threads.  We implement
optimistic locking for LITS to support concurrent threads.

\Paragraph{Datasets}
We use seven real-world string data sets in our experiments, as listed
in Table~\ref{tab:datasets}.
(1) \verb|address| contains 34M addresses in the form of
unit-street-city in the US West~\cite{address}.
%
%
(2) \verb|dblp| contains 7M paper titles in dblp~\cite{dblp}.
(3) \verb|geoname| contains 7M geographical names, such as ``Pic des
Langounelles''~\cite{geoname}.
(4) \verb|imdb| contains 9M actor names in imdb~\cite{imdb}.
(5) \verb|reddit| contains the user names of 26M reddit accounts that
have commented since Dec 2017~\cite{reddit}.
(6) \verb|url| contains 63M urls from the CommonCrawl~\cite{urls}.  An
example is ``\url{http://1000rosanegra.com.ar/index.html}''.
(7) \verb|wiki| contains 43M wiki titles~\cite{wiki}.  An example is
``1980-81\_Mersin\_Idmanyurdu\_season''.

Moreover, we generate four synthetic data sets.
(8) \verb|email| contains 45M synthetic email addresses generated by
the Faker 14.2.1 package using Python 3.6.9.
%
(9) \verb|idcard| contains 63M synthetic Chinese id-card numbers.  A
id-card number is a 18-byte string.  The first 6B represents a region,
such as a city or a county.  The next 8B is the birthday in the form
of ``yyyymmdd''.  Then the remaining 4B assigns a unique code to
distinguish ids with the same 14B prefix.
(10) \verb|phone| contains 50M synthesis phone numbers generated by
the Faker package.
%
(11) \verb|rands| contains 50M randomly generated strings.  The
characters are selected uniformly from \verb|a| to \verb|z|.

For the experiments, all data sets have been processed to remove
duplicate strings, strings containing non-ASCII characters, and
strings longer than 255 characters.  Then, the value for each string
key is a randomly generated 64-bit integer.

\begin{figure*}[t]
	\centering
	$\begin{array}{cc}
			\includegraphics[width=7.7cm]{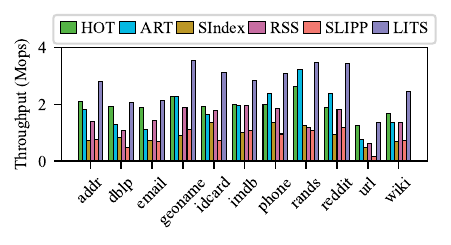} &
			\includegraphics[width=7.7cm]{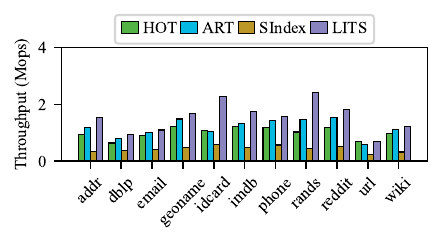}
			\vspace{-4mm}                                     \\
			\mbox{(a) Read-only (YCSB workload C)}          &
			\mbox{(b) Insert-only}
		\end{array}$
	\vspace{-5mm}
	\caption{Index performance for read-only and insert-only workloads.}
	\label{fig:CI}
	\vspace{-0.15in}
\end{figure*}

\begin{figure*}[t]
	\centering
	\vspace{-0.02in}
	\includegraphics[width=15.5cm]{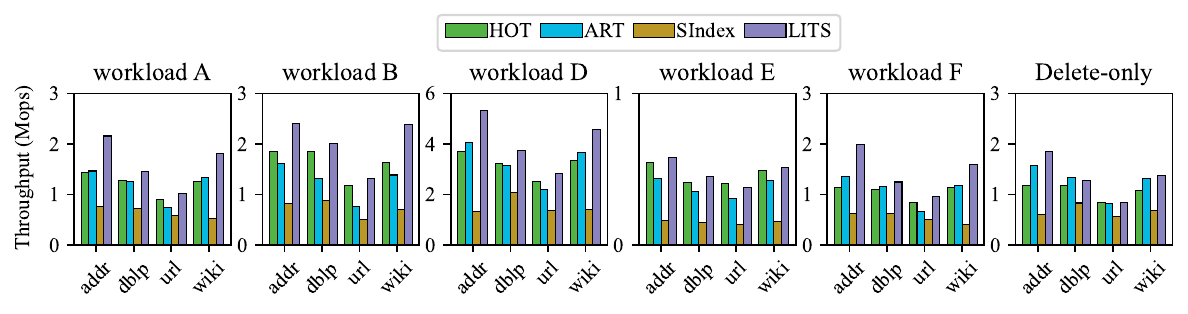}
	\vspace{-0.25in}
	\caption{Index performance for YCSB workloads.}
	\label{fig:BDEF}
	\label{Re:R5D9}
	\vspace{-0.2in}
\end{figure*}

\Paragraph{Workload}
We use six YCSB core workloads: A (50\% read, 50\% update), B (95\%
read, 5\% update), C (100\% read), D (95\% latest-read, 5\% insert), E
(95\% short range scan, 5\% insert), and F (50\% read, 50\%
read-modify-write)~\cite{YCSB}.
For all YCSB workloads except the read-only workload C, we bulkload
the indexes with 80\% of the keys in a data set.  For the read-only
workload, we bulkload 100\% of the keys.  Then, we perform 20M random
operations.  Search keys are randomly selected from the bulkload keys.
Insert keys are randomly selected from the 20\% keys that are new.
Update keys are randomly selected from the entire data set.  For an
existing key, the entry is modified.  For a new key, we will perform an insert operation.
Unless otherwise noted, the random keys are chosen uniformly
at random.  We also perform a set of experiments where the chosen keys follow the zipf distribution with zipf
factor = 1.

Apart from the YCSB workloads, we test insert-only and
delete-only workloads.  For the insert-only workload, we bulkload the
indexes with 50\% keys in a data set, then measure the performance of
randomly inserting all the remaining keys. For the delete-only
workload, we bulkload the indexes with 100\% keys, then measure the
performance of randomly deleting 50\% existing keys.
\label{Re:R2O1_1}

\vspace{-0.1in}


\subsection{Overall Performance}
\label{subsec:perf}

Figure~\ref{fig:CI} and~\ref{fig:BDEF} show the overall index
performance.  We report the results of all 11 data sets for the
read-only (YCSB workload C) and insert-only workloads in
Figure~\ref{fig:CI}.  Due to space limitations, we report the results
of the four largest real-world data sets, i.e., \verb|address|,
\verb|dblp|, \verb|url|, \verb|wiki|, for the YCSB workload A, B, D,
E, F, and the delete-only workload in Figure~\ref{fig:BDEF}.
Experiments on the other data sets show similar trends.



%
%

\begin{figure}[t]
	\centering
	\vspace{-0.05in}
	\includegraphics[width=7.2cm]{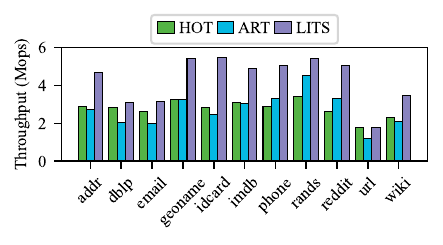}
	\vspace{-0.27in}
	\caption{Index performance for read-only (YCSB workload C) with Zipf distribution.}
	\label{fig:DZ}
	\vspace{-0.22in}
\end{figure}

Figure~\ref{fig:DZ} shows the performance of the read-only
workload under the zipf distribution with zipf factor = 1.
\nottechreport{Due to space limitations, we report more experimental
	results under the zipf distribution in the extended version of the
	paper~\cite{litstr}.}
\techreport{(Please see more experimental results under the zipf
	distribution in Appendix~\ref{app:zipf}.)}  \label{Re:R2O1_2}


From the figures, we see that LITS achieves the best performance for
most workloads and data sets.  (LITS is slightly slower than HOT for
workload E on \verb|url|).  For the read-only workload, LITS achieves
up to 1.93x and 2.23x improvement over HOT and ART, respectively.
Compared to SIndex, LITS demonstrates a performance advantage of
2.26x-3.91x.
LITS also exhibits excellent performance for insert operations.
Compared to HOT, ART, and SIndex, LITS attains up to 2.06x, 2.14x, and
5.31x improvement for the insert-only workload, respectively.
Similarly, LITS achieves up to 2.43x, 2.27x, and 3.99x improvement
over HOT, ART, and SIndex for workload A, B, D, and F.
For the scan-heavy workload E, LITS's performance is comparable with
HOT, and better than ART and SIndex.
Finally, the zipf results show similar trends.
Interestingly, under the zipf distribution, nodes that contain popular
keys tend to stay in the CPU cache, leading to higher index
performance than that with the uniform distribution.

\Paragraph{Index Height} \label{Re:R4O1}
Table~\ref{tab:height} compares the height of different indexes after
bulkloading.
The height of LITS is composed of two parts: LITS (base), which is the
height of the LIT structure including model-based nodes and compact
leaf nodes, and LITS (hot), which is the height of the subtries.
From Table~\ref{tab:height}, we see that the height of LITS is
significantly smaller than HOT, ART, and SLIPP.  This partially
explains the good performance of LITS.  Note that RSS achieves
good tree heights. However, RSS suffers from expensive local search,
and for popular duplicate key prefixes, it has to
visit and compare the string keys.


\begin{table}[t]
  \caption{Comparing the height of index solutions.}
  \label{tab:height}
  \vspace{-0.15in}

  \centering
  \small
  \setlength{\tabcolsep}{2pt}

  \begin{tabular}{|c|c|c|c|c|c|c|c|}
    \hline
    data set & LITS (base) & LITS (hot) & HOT & ART  & SIndex & RSS & SLIPP \\
    \hline\hline
    addr     & 2.7         & 0.2        & 7.0 & 10.2 & 2      & 2.0 & 4.9   \\ \hline
    dblp     & 2.7         & 1.7        & 6.8 & 14.1 & 2      & 2.2 & 7.3   \\ \hline
    url      & 3.0         & 2.1        & 7.8 & 16.1 & 2      & 3.7 & 9.1   \\ \hline
    wiki     & 2.9         & 1.0        & 7.8 & 11.6 & 2      & 2.1 & 5.7   \\ \hline
  \end{tabular}

  \vspace{-0.15in}
\end{table}

\begin{figure}[t]
	\centering
	\includegraphics[width=8cm, viewport=0 0 246 107, clip]{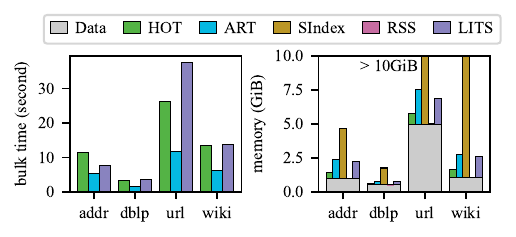}
	\vspace{-0.2in}
	\caption{Bulkload time and memory space consumption}.
	\label{fig:BulkMem}
	\vspace{-0.3in}
\end{figure}

\Paragraph{Bulkload Time}
The left figure in Figure~\ref{fig:BulkMem} compares the bulkload time
of LITS, HOT, and ART.  We bulkload all the keys in each data set.
For HOT and ART, we sort the keys, then insert all the strings into
the index in the sorted order.
From the figure, we see that the bulkload time of LITS is comparable
to that of HOT.


\Paragraph{Space Cost}
The right figure in Figure~\ref{fig:BulkMem} compares the space cost
of LITS, HOT, ART, Sindex, and RSS.  The figure does not
display the space cost of SLIPP; it exceeds 10GB in all four datasets.
The grey bar shows the raw data size.  From the figure, we see that
LITS consumes lower space than ART and SIndex.  SIndex consumes a lot
of space for padding all strings to the maximal length in a data set.
Interestingly, the read-only RSS has the lowest space cost, which is
consistent with the RSS paper~\cite{RSS}.  (Please see the in-depth
analysis of the space consumption of learned indexes in
\techreport{Appendix~\ref{app:space}}
\nottechreport{the extended version of the paper~\cite{litstr}}.)


\label{Re:R5W1_R5D10}

\Paragraph{Scalability} \label{subsec:scalability}
We compare LITS and HOT in scalability experiments.  For each data
set, we bulkload the indexes with 50\% keys in the data set.  Then,
the insert-only workload measures the performance of randomly
inserting the remaining 50\% keys.  After that, the read-only workload
measures the performance of 10M search operations for keys randomly
distributed in the data set.

\begin{figure}[t]
	\centering
	\includegraphics[width=7cm, viewport=0 0 217 132, clip]{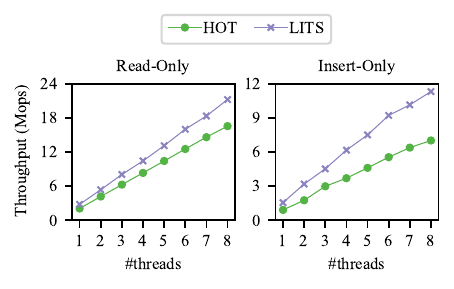}
	\vspace{-0.25in}
	\caption{Scalability on the 34M address dataset.}
	\label{fig:Scala}
	\vspace{-0.2in}
\end{figure}

Figure~\ref{fig:Scala} reports the index throughput for LITS and HOT
varying the number of threads for \verb|address| data
sets.  From the figure, we see that both LITS and HOT achieve nearly
linear scalability.  Compared to HOT, LITS achieves 1.19x -- 1.31x
and 1.52x -- 1.64x improvement for the read-only and insert-only
workloads, respectively.


\subsection{Benefit of HPT}
\label{subsec:HPTPerf}

To understand the benefit of the HPT-based model in LITS, we compare
HPT with existing learned models for strings:
\begin{list}{\labelitemi}{\setlength{\leftmargin}{5mm}\setlength{\itemindent}{-1mm}\setlength{\topsep}{0.5mm}\setlength{\itemsep}{0.5mm}\setlength{\parsep}{0.5mm}}

	\item \emph{Simple Model (SM)}:
	      The simple method to calculate the CDF of a string is to use the
	      equation $x=\frac{c_1}{256} + ... \frac{c_n}{{256}^n}$ to get a
	      monotonic value based on the characters in the string.  SM is used in
	      SLIPP.

	\item \emph{Radix Spline (RS)}:
	      RS is the default CDF model used in Radix String Spline
	      (RSS)~\cite{RSS}.  In each inner node of RSS, a $K$-byte substring of
	      the key string is converted into an integer, and a RS model is used to
	      compute the CDF.  We use the same configuration as the RSS
	      paper~\cite{RSS}.  In the experiments, $K$ is set to 8 and the
	      error-bound in Radix Spline is set to 127.

	\item \emph{SRMI}:
	      SRMI is a string CDF model mentioned in the learned sort
	      paper~\cite{LearnedSort}. SRMI first converts a string into a floating
	      point number using $x=\frac{c_1}{256} + ... + \frac{c_n}{{256}^n}$,
	      then employs a two-layer RMI to compute the CDF from the coded
	      floating point $x$.

\end{list}

\begin{figure}[t]
	\centering
	\includegraphics[width=8cm]{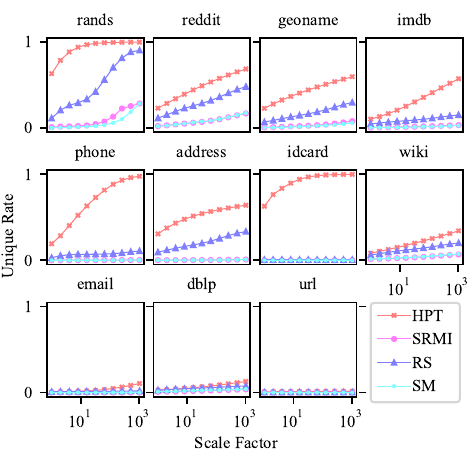}
	\vspace{-0.3in}
	\caption{Unique rate of learned models.}
	\label{fig:UR}
	\vspace{-0.25in}
\end{figure}

\Paragraph{Effectiveness for Distinguishing Strings}
We would like to compare the effectiveness of the learned models for
distinguishing strings in the data sets.  For this purpose, we define
and measure a unique rate metric.

We use a learned model to map a set ${\mathcal{S}}$ of unique strings
to an item array of size $SF\cdot|\mathcal{S}|$, where $SF \geq 1$ is
the scale factor.  In the ideal situation, a perfect learned model
will map every string in ${\mathcal{S}}$ to a separate location in the
array.  However, in the common case, there can be collisions.  That
is, two or more strings are mapped to the same item.  The total number
of occupied item slots after mapping, denoted as $NumValidSlots$, is
always less than or equal to $|\mathcal{S}|$. We define ${UR}_{SF}$
for scale factor $SF$ as follows:
\begin{equation}\label{eqn:ur}
	UR_{SF} = \tfrac{NumValidSlots}{|\mathcal{S}|}
\end{equation}
$UR_{SF}$ is between 0 and 1.  The larger the $UR_{SF}$, the more
effective that the learned model distinguishes keys in the string data
set.

Figure~\ref{fig:UR} shows the unique rates of the four learned models
varying the scale factor from 1 to 1000 for all the 11 data sets.  We
see that HPT achieves the best unique rate for all data sets and under
all the scale factors.  Compared to SM, RS, and SRMI, HPT is more
powerful in distinguishing strings.
For the three data sets with the highest GPKL, i.e., \verb|email|,
\verb|dblp|, and \verb|url|, all the learned models work quite poorly.
These data sets require larger number of bytes to discern one string
from the adjacent string in the sort order, making it hard for the
learned models to separate the strings.

\begin{figure}[t]
	\centering
	\includegraphics[width=7.5cm]{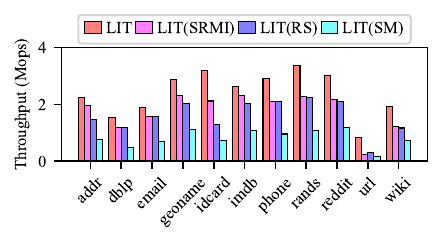}
	\vspace{-0.3in}
	\caption{Index performance with different learned models.}
	\label{fig:ModelCMP}
	\vspace{-0.2in}
\end{figure}

\Paragraph{Index Performance with Different Learned Models}
Figure~\ref{fig:ModelCMP} compares the performance of LIT with
different learned models.  We choose to compare LIT instead of LITS
because the hybrid structure of LITS could mask the performance impact
of the learned model.  The experiments run the read-only workload
(YCSB workload C).
LIT with the HPT model achieves the best index performance.  It is
1.14x -- 3.65x better than the second best, i.e., LIT(SRMI).

\Paragraph{HPT Space and Time Cost}
The HPT is lightweight.  In our experiments, it takes 2MB,
which is orders of magnitude smaller than the data set.  HPT can
easily fit into the CPU cache, and the model prediction using HPT is
fast.  It takes 20--50ns to compute the HPT-based model for an 8-byte
substring.


\subsection{Benefit of Compact Leaf Node}
\label{subsec:CNodePerf}

\begin{figure}[t]
	\centering
	\includegraphics[width=8cm]{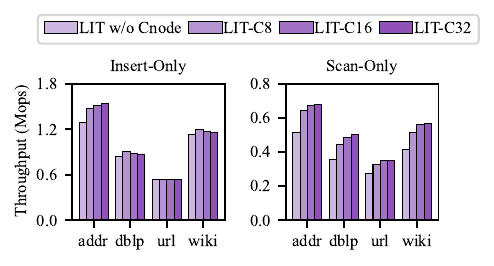}
	\vspace{-0.2in}
	\caption{Comparing LIT with compact node designs.}
	\label{fig:CNodeCMP}
	\vspace{-0.2in}
\end{figure}

We study the benefit of the compact node in this subsection.

\Paragraph{Performance Benefit}
For the same reason as in Section~\ref{subsec:HPTPerf}, we conduct
experiments using LIT rather than LITS.  We compare LIT without
compact nodes, and LIT with compact nodes whose size limit is set to
8, 16, and 32.
Figure~\ref{fig:CNodeCMP} reports the insert-only and scan-only
throughput for the four LIT variants.
From the figure, we see that the introduction of compact leaf nodes
not only improves the performance of the scan operations but also
enhances the performance of the insert operations.  Scan is improved
due to the fact that compact nodes place kv-pointers contiguously,
thereby reducing cache misses for visiting many small nodes during the
scan process.  Insertion is improved because compact nodes tend to
reduce the tree height and avoid extra cache misses caused by visiting
small nodes in deeper levels of the index.

Moreover, we see that the scan throughput increases as the size limit
increases, but becomes relatively flat beyond 16.  The insert
throughput may even suffer when the size limit exceeds 16.  Therefore,
we set the default size limit of compact nodes to 16 in all the other
experiments.

\Paragraph{Variants of Compact Node Implementation}
We consider two variants: 1) pre-allocating 16 entries in compact
nodes; 2) exploiting SIMD for cnode search.
However, our experimental results show that preallocation incurs up to
93\% extra space overhead without significant performance benefits.
The performance improvement with SIMD is less significant since cnode
search is only a small part of the search procedure.
Please see
\techreport{Appendix~\ref{app:cnode}}
\nottechreport{the extended version of the paper~\cite{litstr}}
for more details.

\subsection{Benefit of LIT Enhanced with Subtries}
\label{subsec:HybridPerf}

The LITS mentioned in the above subsections are all LITS-H (i.e., the
hybrid structure of LIT and HOT).
We prefer this combination because the performance characteristics of
LIT and HOT are complementary as shown in Figure~\ref{fig:SDTs}.  HOT
performs well for data sets with large GPKLs and relatively small data
sizes (e.g.  \verb|dblp|).  In comparison, LIT demonstrates better
performance for data sets with small GPKLs and large data sizes
(e.g., \verb|reddit|).

\begin{figure}[t]
	\centering
	\includegraphics[width=7.5cm]{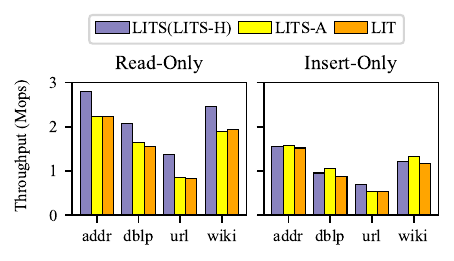}
	\vspace{-0.25in}
	\caption{Comparing LITS-H, LITS-A, and LIT.}
	\label{fig:HybridPerf}
	\vspace{-0.2in}
\end{figure}

In this subsection, we consider the alternative design of combining
LIT and ART, i.e., LITS-A, and LIT without subtries.
Figure~\ref{fig:HybridPerf} compares the read-only and the insert-only
throughput of LITS-H (i.e., LITS), LITS-A, and LIT.
From the figure, we see that LITS-H achieves better performance than
LIT, confirming that the hybrid structure indeed improves index
performance.  For data sets with large GPKLs (e.g., \verb|url|),
LITS-H brings up to 50\% improvement for search performance.
Moreover, compared to LITS-A, LITS-H has higher performance
improvement for search.  For the insert-only workload, LITS-A and
LITS-H show comparable performance.

\Paragraph{GPKL Computation Cost} \label{Re:R5D6}
In the insert-only workload, a single invocation of the GPKL
computation takes 0.8--1.7us.  The total GPKL computation time
contributes to 0.5\%--1.3\% of the total insert time across all data
sets.
%

%
%


\section{Conclusion}
\label{sec:conclusion}

In conclusion, we have presented a novel string index called LITS
(\underline{L}earned \underline{I}ndex with Hash-enhanced Prefix
\underline{T}able and \underline{S}ub-tries).  Our experimental results show
that compared to HOT and ART, LITS achieves up to 2.43x and
2.27x improvement for point operations and comparable scan
performance.




\Paragraph{Acknowledgment}
This work is partially supported by National Key R\&D Program of China
(2023YFB4503600) and Natural Science Foundation of China (62172390).


\bibliographystyle{ACM-Reference-Format}
\bibliography{refs}

\balance

\clearpage{}
\pagenumbering{arabic}

\appendix
\section{appendix}
\label{Appendix}

\subsection{Proof of HPT Accuracy Theorem}
\label{app:hptproof}

We prove Theorem~\ref{thm:hpt} in Section~\ref{subsec:HPT}.

\begin{proof}

    Suppose $n_{\mathcal{P}c}$ is the number of occurrences of prefix
    $\mathcal{P}$ followed by character $c$.  Then,
    $prob(c|\mathcal{P})=\tfrac{n_{\mathcal{P}c}}{n_{\mathcal{P}}}$.

    Suppose for any other prefix hashed to the HPT[$hash({\mathcal{P}})$]
    row, there are in total $d_c$ occurrences of such a prefix followed by
    character $c$, where $0 \leq d_c \leq d$.  Then,
    $HPT[hash({\mathcal{P}})][c].prob = \tfrac{n_{\mathcal{P}c} +
            d_c}{n_{\mathcal{P}} + d}$.

    Hence, we have the following:
    \begin{equation}\label{eqn:hpt_abs_err}
        \begin{aligned}
             & HPT[hash({\mathcal{P}})][c].prob - prob(c|{\mathcal{P}}) =
            \frac{n_{\mathcal{P}c} + d_c}{n_{\mathcal{P}} + d} - \frac{n_{\mathcal{P}c}}{n_{\mathcal{P}}}       \\
             & \leq \frac{n_{\mathcal{P}c} + d}{n_{\mathcal{P}} + d} - \frac{n_{\mathcal{P}c}}{n_{\mathcal{P}}}
            = \frac{d(n_{\mathcal{P}} - n_{\mathcal{P}c})}{n_{\mathcal{P}}(n_{\mathcal{P}}+ d)}
            = \frac{1 - \tfrac{n_{\mathcal{P}c}}{n_{\mathcal{P}}}}{\tfrac{n_{\mathcal{P}}}{d}+1}
            \leq \frac{1}{\tfrac{n_{{\mathcal{P}}}}{d}+1}
        \end{aligned}
    \end{equation}

    \begin{equation}\label{eqn:hpt_abs_err2}
        \begin{aligned}
             & prob(c|{\mathcal{P}}) - HPT[hash({\mathcal{P}})][c].prob =
            \frac{n_{\mathcal{P}c}}{n_{\mathcal{P}}} - \frac{n_{\mathcal{P}c} + d_c}{n_{\mathcal{P}}+d}   \\
             & \leq \frac{n_{\mathcal{P}c}}{n_{\mathcal{P}}} - \frac{n_{\mathcal{P}c}}{n_{\mathcal{P}}+d}
            = \frac{d \cdot n_{\mathcal{P}c}}{n_{\mathcal{P}}(n_{\mathcal{P}}+ d)}
            \leq \frac{d}{n_{\mathcal{P}}+ d}
            = \frac{1}{\tfrac{n_{{\mathcal{P}}}}{d}+1}
        \end{aligned}
    \end{equation}

    From the above, it is clear that the absolute error is at most
    $\tfrac{1}{\tfrac{n_{{\mathcal{P}}}}{d}+1}$.

\end{proof}

\subsection{Index Search and Insert Algorithms}
\label{app:algo}

\newcommand{\code}[1]{\texttt{\small #1}}

Section~\ref{subsec:overview} has described the search, insert,
delete, update, and scan index operations.  To make the description
more concrete, we present the pseudo-code of the index search and
index insert algorithms in this subsection.

\Paragraph{Index Search Algorithm}
Algorithm~\ref{alg:LITS_Search} lists the procedures for searching a
string key $s$ in LITS.
The starting point is the root node (Line 2).  LITS goes into a while
loop (Line 3).  A loop iteration visits a node at one level of the
tree.  At each level of the tree, LITS examines the type of the item
(Line 4--10).  As depicted previously in Figure~\ref{fig:lits}, the
node type is a 3-bit field in a 64-bit item.

First, if the item is a pointer pointing to a sub-trie (e.g.  HOT),
LITS calls the sub-trie's search function (Line 4--5).

Second, If the item is a pointer pointing to a single kv-entry, LITS
invokes \code{singleVerify} to verify the key and return the value if
the key matches the search key (line 6--7).

Third, if the item is a pointer pointing to a compact leaf node, LITS
calls \code{compactSearch} (Line 8--9).   \code{compactSearch}
computes the hash value of the search key (line 22), then compares the
hash value with the hash codes stored in the h-pointers (Line 23--24).
It dereferences an h-pointer only if the hash value of the search key
matches the hash code in the h-pointer (Line 25--26).

Fourth, if the item is an empty slot, LITS returns NULL.  The NULL
return value indicates that the search does not find any matching
index entry (Line 10--11).

Finally, if none of the above four conditions are met, the item must
be a pointer pointing to a model-based node.  Under this circumstance,
LITS calls the \code{locate} procedure to locate the item in the next
level of the tree.  It compares the prefix recorded in model-based
node with the search key (line 31--34).  If the prefix does not match,
LITS returns the first (the last) item in the item array for the case
where the search key prefix is less (greater) than the recorded prefix
(line 32--34).  If the prefix matches, LITS locates the item using
HPT-based learned model (Line 35--37).

\begin{algorithm}[t]
    \caption{Index search algorithm in LITS.}
    \label{alg:LITS_Search}
    \begin{algorithmic}[1]
        \Procedure{LITSSearch}{string s}
        \State item = root
        \While{true}
        \If{item.type == Trie}
        \State \textbf{return} trieSearch(item.trie(), s)
        \EndIf
        \If{item.type == SingleEntry}
        \State \textbf{return} singleVerify(item.entry(), s)
        \EndIf
        \If{item.type == CompactLeaf}
        \State \textbf{return} compactSearch(item.cnode(), s)
        \EndIf
        \If{item.type == EmptySlot}
        \State \textbf{return} NULL
        \EndIf
        \State item = locate(item.modelNode(), s, item.prefixLen())
        \EndWhile
        \State \textbf{return} NULL
        \EndProcedure
        \State
        \Procedure{singleVerify}{singleEntry entry, string s}
        \If{strcmp(entry.key, s) == 0}
        \State \textbf{return} entry.value()
        \Else
        \State \textbf{return} NULL
        \EndIf
        \EndProcedure
        \State
        \Procedure{compactSearch}{compactLeaf cnode, string s}
        \State hashVal = hash(s)
        \For{$entry$ in $cnode$}
        \If{entry.hashCode == hashVal}
        \If{entry.key == s}
        \State \textbf{return} entry.value()
        \EndIf
        \EndIf
        \EndFor
        \State \textbf{return} NULL
        \EndProcedure
        \State
        \Procedure{locate}{modelNode node, string s, int prefixlen}
        \State len = node.getLen()
        \If{strncmp(s, node.prefix, prefixlen) < 0}
        \State \textbf{return} node.itemArray[0]
        \EndIf
        \If{strncmp(s, node.prefix, prefixlen) > 0}
        \State \textbf{return} node.itemArray[len-1]
        \EndIf
        \State cdf = hpt.getCDF(s + prefixLen)
        \State cdf = node.k * cdf + node.b
        \State pos = max(1, min(len-2, cdf * len))
        \State \textbf{return} node.itemArray[pos]
        \EndProcedure
    \end{algorithmic}


\end{algorithm}

\begin{algorithm}[t]
    \caption{Index insert algorithm in LITS.}
    \label{alg:LITS_Insert}
    \begin{algorithmic}[1]
        \Procedure{LITSInsert}{string s, int val}
        \State item = root
        \State result = false
        \State path = []
        \While{true}
        \If{item.type == Trie}
        \State result = trieInsert(item.trie(), s, val)
        \State \textbf{break}
        \EndIf
        \If{item.type == SingleEntry}
        \State result = singleInsert(item, s, val)
        \State \textbf{break}
        \EndIf
        \If{item.type == CompactLeaf}
        \State result = compactInsert(item, s, val)
        \State \textbf{break}
        \EndIf
        \If{item.type == EmptySlot}
        \State item = setEntry(s, val)
        \State result = true
        \State \textbf{break}
        \EndIf
        \State recordPath(path, item) \Comment{record the path}
        \State item = locate(item.modelNode(), s)
        \EndWhile
        \If{result == true}
        \State incCount(path)
        \EndIf
        \State \textbf{return} result
        \EndProcedure
        \State
        \Procedure{singleInsert}{Item\& item, string s, int val}
        \If{strcmp(item.entry().key, s) == 0} \Comment{key exists}
        \State \textbf{return} false
        \Else
        \State item = createCnode(entry, s, val)
        \State \textbf{return} true
        \EndIf
        \EndProcedure
        \State
        \Procedure{compactInsert}{Item\& item, string s, int val}
        \State cnode = item.cnode()
        \State pos = BinarySearch(cnode, s)
        \If{pos == -1} \Comment{key exists}
        \State \textbf{return} false
        \EndIf
        \If{cnode.keyCount() < CnodeCapacity}
        \State item = cnode.insert(s, val, pos)
        \Else
        \State item = PMSSBuild(cnode, s, val)
        \EndIf
        \State \textbf{return} true
        \EndProcedure
        \State
        \Procedure{incCount}{Path path}
        \For{$item$ in $path$}
        \State node = item.modelNode()
        \State node.keyCount += 1
        \If{node.keyCount >= 2 * node.getLen()}
        \State item = PMSSBuild(node)
        \State \textbf{return}
        \EndIf
        \EndFor
        \State \textbf{return}
        \EndProcedure
    \end{algorithmic}
\end{algorithm}

\Paragraph{Index Insert Algorithm}
Algorithm~\ref{alg:LITS_Insert} lists the procedures for inserting a
string key $s$ and its value $val$ in LITS.  Similar to the search
algorithm, LITS starts from the root node (Line 2), and searches the
insert key using a while loop (Line 5--20).   A loop iteration visits
a node at one level of the tree.  For a model-based node, like the
search algorithm, LITS calls the \code{locate} procedure to locate the
item in the next level of the tree (Line 20).  If the item satisfies
one of the if-conditions (Line 6--18), then LITS processes the
insertion by invoking relevant procedures and exits from the while
loop.

First, if the item points to a sub-trie, LITS calls the sub-trie's
insert function (Line 6--8).

Second, If the item points to a single kv-entry, LITS invokes
\code{singleInsert} (Line 9--11).   It checks the insert key against
the stored key (Line 26).  If there is a match, then the key already
exists.  LITS returns false.  Otherwise, LITS creates a new compact
node with two index entries, and updates the item (Line 29).

Third, if the item points to a compact leaf node, LITS calls
\code{compactInsert} (Line 12--14).   The keys in a compact node are
sorted for better scan performance.  Hence, \code{compactInsert}
performs a binary search to locate the position to insert $s$ (Line
34).  If the key exists in the compact node, then LITS returns false
(Line 35--36).  Otherwise, LITS inserts the new key into the cnode.
The insertion actually creates a new cnode, copies the existing keys
and the new key to the new node, and updates the item (Line 38).  If
the compact node is full, LITS invokes the PMSS-based decision and
creates either a model-based node or a sub-trie (Line 40).

Fourth, if the item is an empty slot, LITS stores the pointer to the
kv-entry in the item (Line 15--18).

LITS records the path of the model-based nodes from the root to the
leaf (Line 19).  After the while loop, LITS increases the key count in
each model-based node on the path (Line 22).  The \code{incCount}
procedure also checks if the key count exceeds twice of the item array
length in a model-based node (Line 47).  If this is the case, it
invokes a resizing procedure (Line 48).


\subsection{Complexity Analysis of LITS}
\label{app:complexity}

The time and space cost of tries (e.g., ART~\cite{ART} and
HOT~\cite{HOT}) have been extensively measured and studied.  In the
following, we mainly focus on the time and space cost of LITS after
removing the subtries.

\Paragraph{Tree Height}
In each inner node (i.e., model-based node), we consider the fraction
of keys mapped to each index slot.  LITS detects the case where over
50\% of the keys are mapped to an index slot.  In such a case, LITS
employs a subtrie for the child node corresponding to the index slot.
Consequently, it guarantees that the subtree rooted at a (non-subtrie)
node contains less than 50\% of the keys of the subtree rooted at its
parent model-based node.  As a result, the height of the non-subtrie
part of LITS is $O(logN)$ in the worst case, where $N$ is the number
of keys in the index.

\Paragraph{Search Cost}
A search in LITS moves from the root to a leaf node.  The cost depends
mainly on the height of the tree.  Therefore, the search cost for
the non-subtrie part of LITS is $O(logN)$.

\Paragraph{Insert/Delete/Update Cost}
Compared to the search cost, the insert cost is more complex.  We need
to consider the possibility of node adjustment.   Similar to
LIPP~\cite{LIPP}, we can show that the amortized cost for insert
operations is $O({log}^2N)$.

First, we consider the cost of adjusting a node $A$ with $N$ index
entries.  We need to read/copy all the keys, and compute the GPKL.
This costs $O(N)$.  If $A$ is a root of a subtree with $N$ index
entries, in the worst case, each level in the subtree could cost
$O(N)$.  Hence, the total adjustment cost can be up to $O(NlogN)$.

Second, we perform amortized analysis.  The idea is to save extra
$O(logN)$ credits at each node along the traversal path of insert
operations.  Then adjustment operations consume the saved credits.  In
this way, we can show that the amortized cost for an insert is
$O({log}^2N)$.

For the delete cost, we can use a similar argument as the insert cost
to show that the amortized cost for a delete is $O({log}^2N)$.

An update of a value for an existing key performs a search for the
key, then an update for the value.  The complexity is $O(logN)$.

\begin{figure*}[t]
    \centering
    $\begin{array}{ccccc}
                                                                         &
            \mbox{\small\bf \hspace{1mm} YCSB-A (50\% read 50\% update)} &
            \mbox{\small\bf \hspace{1mm} YCSB-B (95\% read 5\% update)}  &
            \mbox{\small\bf \hspace{2mm} YCSB-C (100\% read)}            &
            \mbox{\small\bf \hspace{1mm} YCSB-F (50\% read 50\% rmw)}
            \\
            \rotatebox{90}{\small\bf \hspace{7mm} Intel Core i7-4770}    &
            \includegraphics[width=4cm]{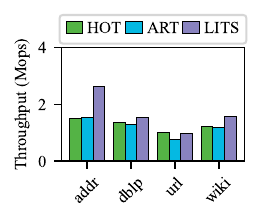}       &
            \includegraphics[width=4cm]{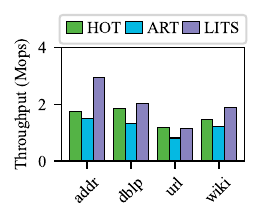}       &
            \includegraphics[width=4cm]{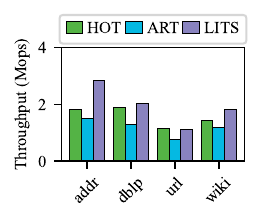}       &
            \includegraphics[width=4cm]{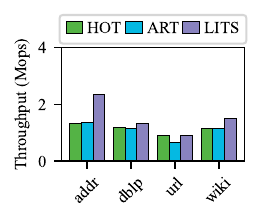}
            \vspace{-2mm}
            \\
            \rotatebox{90}{\small\bf \hspace{7mm} Intel Core i7-9700}    &
            \includegraphics[width=4cm]{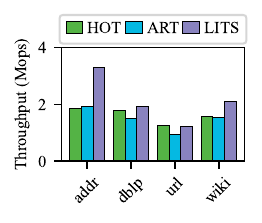}       &
            \includegraphics[width=4cm]{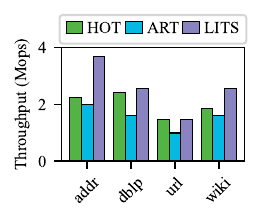}       &
            \includegraphics[width=4cm]{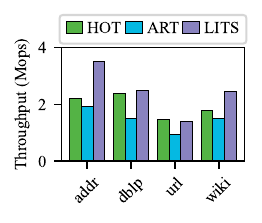}       &
            \includegraphics[width=4cm]{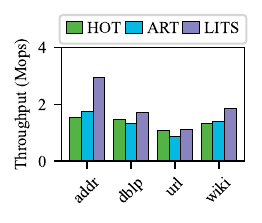}
            \vspace{-2mm}
            \\
            \rotatebox{90}{\small\bf Intel Xeon Platinum 8380}           &
            \includegraphics[width=4cm]{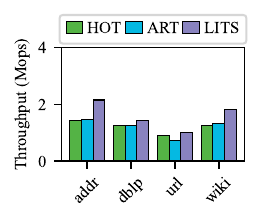}        &
            \includegraphics[width=4cm]{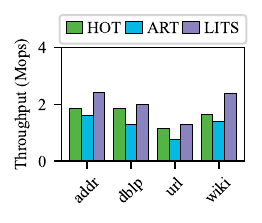}        &
            \includegraphics[width=4cm]{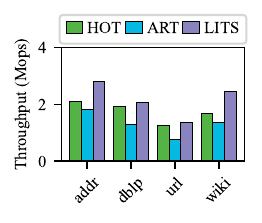}        &
            \includegraphics[width=4cm]{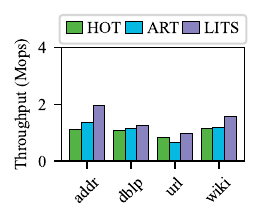}
            \vspace{-2mm}
        \end{array}$
    \vspace{-5mm}
    \caption{Index performance on three different hardware platforms.}
    \label{fig:YCSB_machine}
    \vspace{-0.1in}
\end{figure*}


\begin{figure*}[t]
    \centering
    $\begin{array}{cc}
            \includegraphics[width=8cm]{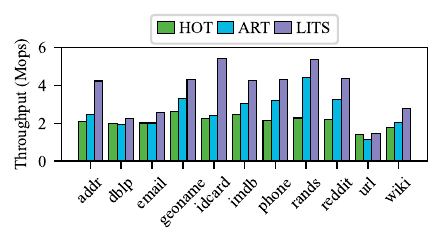} &
            \includegraphics[width=8cm]{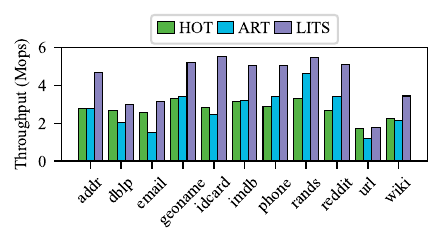}
            \vspace{-3mm}                                          \\
            \mbox{(a) YCSB-A}                                    &
            \mbox{(b) YCSB-B}                                      \\
            \includegraphics[width=8cm]{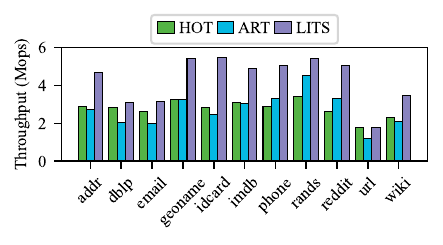} &
            \includegraphics[width=8cm]{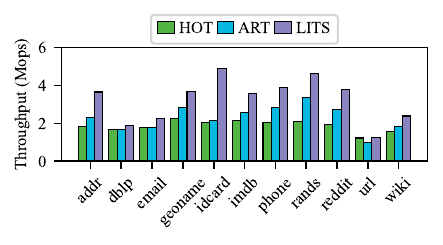}
            \vspace{-3mm}                                          \\
            \mbox{(c) YCSB-C}                                    &
            \mbox{(d) YCSB-F}
        \end{array}$
    \vspace{-4mm}
    \caption{Index performance for YCSB A, B, C and F workloads. (Zipf Distribution)}
    \label{fig:zipf}
    \vspace{-0.2in}
\end{figure*}

\Paragraph{Space Cost}
First, the HPT is a fixed sized data structure.  Second, for every
model-based node or compact leaf node, the size of the node is
proportional to the number of index entries that the node contains.
Therefore, the space cost of a node with $N$ index entries is $O(N)$.
Since the tree height is $O(logN)$, the space cost of the tree is
$O(NlogN)$.

\subsection{Performance Across Different Platforms}
\label{app:platform}

We compare LITS, HOT, and ART on three hardware platforms.  As
described in Section~\ref{subsec:setup}, the default experimental
machine is equipped with Intel Xeon Platinum 8380 CPUs.  In addition
to the default configuration, we also run experiments on two different
machines in this subsection:
\begin{list}{\labelitemi}{\setlength{\leftmargin}{5mm}\setlength{\itemindent}{-1mm}\setlength{\topsep}{0.5mm}\setlength{\itemsep}{0.5mm}\setlength{\parsep}{0.5mm}}

    \item \emph{Intel Core i7-4770}:
          The machine is equipped with a 3.9 GHz (turbo frequency) Intel Core
          i7-4770 (32KB L1 cache, 256 KB L2 cache, and 8 MB L3 cache).

    \item \emph{Intel Core i7-9700}:
          The machine is equipped with a 4.7 GHz (turbo frequency) Intel Core i7-9700
          (256KB L1 cache, 2 MB L2 cache, and a 12 MB L3 cache).

\end{list}

\begin{figure}[t]
    \centering
    \includegraphics[width=8cm]{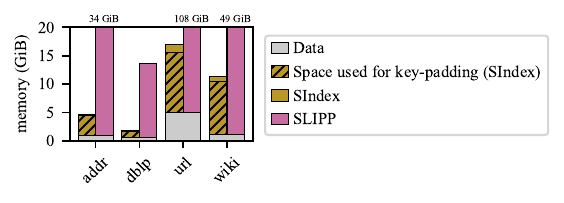}
    \vspace{-0.2in}
    \caption{Space used by SIndex and SLIPP.}
    \label{fig:DetailSpace}
    \label{Re:DS}
    \vspace{-0.2in}
\end{figure}

Figure~\ref{fig:YCSB_machine} compares four YCSB workloads on three
different hardware platforms.  Each row of sub-figures shows the
performance on a hardware platform.  Each column of sub-figures shows
the performance of a workload.
From the figure, we see that LITS achieves the best performance for
all the four workloads across the three hardware platforms.

\subsection{Performance for Zipf Distribution}
\label{app:zipf}

Figure~\ref{fig:zipf} shows the performance of YCSB A, B, C and F
workloads under the zipf distribution with zipf factor = 1.  Note that
workloads with insert operations would incur many redundant and thus
invalid insertions under the zipf distribution.  Such inserts would be
ignored.  Since this situation is rare in practice, we do not show
workloads with insertions.
Similarly, repeatedly deleting the popular keys does not make much
sense.  Therefore, we do not show results for the delete-only
workload.

\balance  

\subsection{In-depth Analysis of Space Consumption of Learned Indexes
    for Strings}
\label{app:space}

We study the space consumption of SIndex, RSS, SLIPP, and LITS in this
subsection.
\begin{list}{\labelitemi}{\setlength{\leftmargin}{5mm}\setlength{\itemindent}{-1mm}\setlength{\topsep}{0.5mm}\setlength{\itemsep}{0.5mm}\setlength{\parsep}{0.5mm}}

    \item \emph{SIndex}:
          SIndex requires all strings to be padded to the maximal length in a
          string data set.  As shown in Figure~\ref{fig:DetailSpace}, the key
          padding significantly increases the space overhead.

    \item \emph{RSS}:
          Each RSS inner node has a Radix Spline.  The Radix Spline is about 1MB
          large using 18 bits near the root (based on the default configuration
          of the Radix Spline code).  We follow the RSS paper to use 6 bits in
          the Radix Spline near the leaves.  Since RSS is read only, it can use
          an array index to access the key-value entries in the data array.
          This reduces the space cost of RSS.  RSS has the lowest space cost
          among all compared index solutions, which is consistent with the RSS
          paper~\cite{RSS}.


    \item \emph{SLIPP}:
          LIPP (SLIPP) employs an aggressive node allocation strategy.  For a
          node with $m$ elements, where $m < 100K$, it allocates an item array
          with $6m$ slots.  This incurs significant memory overhead for small to
          medium sized nodes. In comparison, we set the maximum item array size
          to be up to 2x of the number of elements to save space in LITS.
          Moreover, SLIPP is much higher than LITS because its learned models
          are less effective in distinguishing string key prefixes.  This also
          contributes to the large space overhead of SLIPP.

    \item \emph{LITS}:
          The following factors contribute to the modest space overhead of LITS.
          First, the tree height of LITS is low.  The HPT model can more
          effectively distinguish different string keys.  The compact leaf nodes
          compact subtrees that contain a small number of elements.
          Second, we reduce the space for each item by exploiting the upper 16
          bits of the pointer to store meta information, as shown in
          Figure~\ref{fig:lits} in Section~\ref{subsec:overview}.

\end{list}

\subsection{Variants of Compact Leaf Node}
\label{app:cnode}

We consider two factors in the implementation of compact leaf nodes:
pre-allocation vs. no-preallocation (default), and SIMD vs. no SIMD
(default).

\Paragraph{Pre-allocation vs. No Pre-allocation}
We compare the two methods for supporting inserts in compact nodes as
described in Section~\ref{subsec:cnode}.
Table~\ref{tab:prealloc} compares both the space cost and the insert
throughput of the two methods: pre-allocating 16 entries in compact
nodes with reserved empty slots vs. no pre-allocation.  The latter is
the default method in LITS.
From the table, we see that preallocation incurs up to 93\% additional
space overhead due to the reserved empty slots.
This impact is particularly pronounced on data sets with lower GPKLs,
as LITS tends to have fewer sub-tries and more compact nodes.
On the other hand, preallocation does not show significant improvement
of insert throughput.
As a result, we choose the no-pre-allocation method for compact nodes
in LITS to avoid the significant space overhead.







\Paragraph{SIMD vs. No SIMD}
We attempt to exploit SIMD instructions to accelerate the matching of
hash codes in cnode.  Specifically, we employ AVX512 to
simultaneously check eight 8-byte h-pointers.  We match the 16-bit
hash codes of the h-pointers against the hash code of the search key.
However, our experiments show that the improvement of search
performance with SIMD is less than 1\% across various data sets.  This
is because cnode search is only a small part of the overall search
procedure.  LITS visits model-based nodes from the root node to the
parent of the cnode before reaching the cnode.  After locating a
matching hash code, LITS compares the search key and the stored string
key to verify the match.  Both the tree traversal and the string key
comparison are often more time consuming than the search within the
cnode.








\begin{table}[t]

    \caption{Space cost and insert throughput of pre-allocation
        normalized to that of no pre-allocation.}
    \label{tab:prealloc}
    \vspace{-0.15in}

    \small
    \setlength{\tabcolsep}{6pt}

    \begin{tabular}{|l|ccccc|}
        \hline
        Data set          & addr          & idcards       & email & geoname & phones        \\
        \hline\hline
        space cost        & \textbf{1.49} & \textbf{1.93} & 1.01  & 1.33    & \textbf{1.58} \\
        \hline
        insert throughput & 1.006         & 1.000         & 1.009 & 1.000   & 1.019         \\
        \hline
    \end{tabular}

    \vspace{0.05in}
    \setlength{\tabcolsep}{5.5pt}

    \begin{tabular}{|l|cccccc|}
        \hline
        Data set          & dblp  & imdb  & rands         & reddit & url   & wiki  \\
        \hline\hline
        space cost        & 1.09  & 1.17  & \textbf{1.66} & 1.32   & 1.01  & 1.12  \\
        \hline
        insert throughput & 1.010 & 1.011 & 0.992         & 1.016  & 1.014 & 1.000 \\
        \hline
    \end{tabular}

    \vspace{-0.15in}
\end{table}

\end{document}